\documentclass[12pt,onecolumn]{IEEEtran}

\usepackage{graphicx,amssymb,amsmath}
\usepackage{multicol}
\usepackage[noadjust]{cite}
\usepackage{setspace}               
\usepackage{stfloats}

\usepackage{amsthm}
\usepackage{amsmath}
\usepackage{flushend}
\usepackage{cases,subeqnarray}
\usepackage{bm,multirow,bigstrut}
\usepackage{textcomp}
\usepackage{latexsym,bm}
\usepackage{booktabs,changebar}
\usepackage{xcolor}
\usepackage{mathtools}
\usepackage{dsfont}
\usepackage{extarrows}
\usepackage{mathrsfs}
\usepackage{cite}
\usepackage{bm}
\usepackage{cleveref}
\usepackage{multicol}       
\usepackage{multirow}       
\usepackage{array}          
\definecolor{crimson}{RGB}{192,0,0}         
\definecolor{navy}{RGB}{47,85,151}         

\makeatletter                               
\newif\if@restonecol
\makeatother

\makeatletter
\newif\if@restonecol
\makeatother

\usepackage[linesnumbered,ruled,vlined]{algorithm2e}
\usepackage{algpseudocode}
\usepackage{amsmath}
\renewcommand{\arraystretch}{1.5} %

\theoremstyle{plain}

\newtheorem{lemm}{Lemma}

\newtheorem{coro}{Corollary}
\newtheorem{assu}{Assumption}
\theoremstyle{plain}

\IEEEoverridecommandlockouts

\begin{document}

\title{Structured Massive Access for Scalable Cell-Free Massive MIMO Systems}

\author{Shuaifei~Chen,~\IEEEmembership{Student Member,~IEEE}, Jiayi~Zhang,~\IEEEmembership{Member,~IEEE}, Emil~Bj{\"o}rnson,~\IEEEmembership{Senior Member,~IEEE}, Jing Zhang, and Bo Ai,~\IEEEmembership{Senior Member,~IEEE}
\thanks{S. Chen, J. Zhang and J. Zhang are with the School of Electronic and Information Engineering, Beijing Jiaotong University, Beijing 100044, China. (e-mail: jiayizhang@bjtu.edu.cn).}
\thanks{E. Bj\"{o}rnson is with the Department of Electrical Engineering (ISY), Link\"{o}ping University, SE- 58183 Link\"{o}ping, Sweden. (e-mail: emil.bjornson@liu.se).}
\thanks{B. Ai is with State Key Laboratory of Rail Traffic Control and Safety, Beijing Jiaotong University, Beijing 100044, China. (e-mail: boai@bjtu.edu.cn).}
}
\maketitle

\begin{abstract}
How to meet the demand for increasing number of users, higher data rates, and stringent quality-of-service (QoS) in the beyond fifth-generation (B5G) networks?
Cell-free massive multiple-input multiple-output (MIMO) is considered as a promising solution, in which many wireless access points cooperate to jointly serve the users by exploiting coherent signal processing.
However, there are still many unsolved practical issues in cell-free massive MIMO systems, whereof scalable massive access implementation is one of the most vital.
In this paper, we propose a new framework for structured massive access in cell-free massive MIMO systems, which comprises one initial access algorithm, a partial large-scale fading decoding (P-LSFD) strategy, two pilot assignment schemes, and one fractional power control policy.
New closed-form spectral efficiency (SE) expressions with maximum ratio (MR) combining are derived.
The simulation results show that our proposed framework provides high SE when using local partial minimum mean-square error (LP-MMSE) and MR combining.
Specifically, the proposed initial access algorithm and pilot assignment schemes outperform their corresponding benchmarks, P-LSFD achieves scalability with a negligible performance loss compared to the conventional optimal large-scale fading decoding (LSFD), and scalable fractional power control provides a controllable trade-off between user fairness and the average SE.
\end{abstract}

\begin{IEEEkeywords}
Beyond 5G network, cell-free massive MIMO, massive access, AP selection, pilot assignment, user-centric network.
\end{IEEEkeywords}

\IEEEpeerreviewmaketitle

\section{Introduction}

Cellular massive multiple-input multiple-output (MIMO) is recognized as a component of the fifth-generation (5G) networks \cite{marzetta2010noncooperative,larsson2014massive,wong2017key,andrews2014will,parkvall2017nr}. Looking into the future, beyond 5G networks are expected to handle a significantly larger number of accessing users and deliver high data rates, while providing a more uniform quality-of-service (QoS) throughout the entire network \cite{zhang2019multiple}. These goals can be potentially be achieved by cell-free massive MIMO \cite{ngo2017cell,nayebi2017precoding,interdonato2019ubiquitous,zhang2019cell}, which inherits several virtues from cellular massive MIMO (in particularly \emph{favorable propagation}) while being capable of reaching the beyond 5G requirements.

The basic idea of cell-free massive MIMO is to deploy a large number of access points (APs), which are arbitrarily distributed in the coverage area and connected to a central processing unit (CPU).
Under the coordination and computational assistance from the CPU, the APs jointly serve all user equipments (UEs) on the same time-frequency resource by coherent joint transmission and reception \cite{ngo2017total,bjornson2019making,zhang2018performance}. Hence, cell-free massive MIMO can be viewed as a structured approach to massive access.
Firstly, its macro-diversity can greatly improve the coverage probability compared to cellular technology \cite{ngo2017cell,nayebi2017precoding,bjornson2019making}.
Secondly, interference is managed by letting a user-centric subset of the APs serve each user \cite{buzzi2017cell,interdonato2019scalability,bjornson2019scalable}.
These two features allow cell-free massive MIMO to accommodate more UEs than cellular networks, where inter-cell interference and pilot shortage are the limiting factors.

Channel state information (CSI) is essential in multiple antenna systems, both cellular and cell-free \cite{bjornson2017massive}. It is usually acquired through pilot transmission between the UEs and APs. The pilot resources are limited due to the natural channel variations in time and frequency domain, thus pilots must be reused between UEs in cell-free massive MIMO  \cite{ngo2017cell,nayebi2017precoding,interdonato2019ubiquitous}, leading to the so-called \emph{pilot contamination}.
This phenomenon both reduces the channel estimation quality, which makes coherent transmission less effective, and makes it harder to reject interference between pilot-sharing UEs \cite{bjornson2017massive}. To limit these negative effects, a proper pilot assignment is critical in cell-free massive MIMO networks, particularly in a massive access scenario when the number of UEs $K$ is roughly the same as the number of APs $L$.


While the benefits of cell-free massive MIMO over cellular massive MIMO are well established, it will be very challenging to achieve a practically feasible implementation architecture.
The first steps toward a scalable implementation are taken in \cite{interdonato2019scalability,bjornson2019scalable}, where the authors declare that a cell-free massive MIMO network is required to guarantee the complexity and resource requirements of signal processing
to be finite for each AP as $K \to \infty $.
Although an algorithm for joint initial access, pilot assignment, and power control in cell-free massive MIMO networks have been proposed in \cite{bjornson2019scalable}, it was not designed for massive access scenario with $L \approx K$ and won't perform well in this case.
Hence, the main objective of this paper is to design a framework for structured massive access in scalable cell-free massive MIMO networks, including initial access, data decoding, pilot assignment, and power control. The imperfect CSI, spectral efficiency (SE), user density, and fairness among the UEs are also taken into account.

\subsection{Related Works}
There is a large body of research on massive access in cellular massive MIMO  \cite{de2017random,sun2015beam,lin2017new,de2017randomp,sorensen2018coded,fengler2019massive,8125754,8937497}.
According to the user density in the network, massive access can be divided into \emph{structured access} and \emph{random access}.
When the number of pilots is smaller than the number of UEs, but not dramatically like in Internet of Things (IoT) networks \cite{wang2020wirelessly}, structured access where each user is allocated a dedicated pilot resource is preferable \cite{de2017random}.
In contrast, random access might outperform structured access in highly crowded scenarios.
Structured access has been considered in \cite{sun2015beam,lin2017new}.
Specifically, the authors in \cite{sun2015beam} proposed a beam division multiple access to simultaneously serve multiple UEs via different beams in a multiuser massive MIMO network. From the perspective of array signal processing, the authors in \cite{lin2017new} treated the multiuser massive MIMO as a type of non-orthogonal angle division multiple access to simultaneously serve multiple UEs.
On the other hand, in \cite{de2017randomp}, the authors improved the random access performance by averaging the pilot contamination across the transmission slots.
In \cite{sorensen2018coded}, the authors viewed the contaminated pilot signals as a graph code and analytically optimized performance by performing iterative belief propagation.
The authors in \cite{fengler2019massive} proposed a non-Bayesian algorithm to detect the activity of a large number of UEs for massive unsourced random access.
Since the cell-free massive MIMO is widely used in indoor and hotspots scenarios, we focus on improving the structured access methods by suppressing the pilot contamination.

Cell-free massive MIMO was proposed in \cite{ngo2017cell,nayebi2017precoding}, but builds on the heritage of coordinated multipoint \cite[Sec.~7.4.3]{bjornson2017massive}.
Four different ways to divide the signal processing between the APs and CPU are considered in \cite{bjornson2019making}. The most promising distributed implementation uses minimum mean-squared-error (MMSE) combining along with large-scale fading decoding (LSFD) \cite{nayebi2016performance}.
While all APs initially served all UEs, the user-centric approach has later become the leading way to achieve a practically implementable architecture \cite{buzzi2017cell,interdonato2019scalability,bjornson2019scalable,rezaei2020underlaid}.
Several pilot assignment methods have been considered in the literature \cite{interdonato2019ubiquitous}, including random assignment and brute-force optimization. A greedy algorithm was considered in \cite{ngo2017cell} but it focused on limiting the coherent interference, which might not be the dominant part of pilot contamination and is also not scalable.
Additionally, pilot assignment schemes based on tabu-search and K-means clustering were provided in \cite{liu2019tabu} and  \cite{attarifar2018random}, respectively.
The former is also not scalable, while we will look into ways to improve the {K-means approach} in this paper. It was shown in \cite{bjornson2019scalable} that pilot assignment can be made scalable by providing each accessing UE with the least bad pilot, but no optimization was carried out and the method is only evaluated for $L \gg K$ for which pilot assignment is fairly easy.

\subsection{Main Contributions}
In this paper, we design a structured massive access uplink framework for scalable cell-free massive MIMO systems. Our main contributions are given as follows.
{
\begin{enumerate}
  \item
  We propose a scalable partial LSFD (P-LSFD) strategy for multi-antenna APs, which achieves roughly the same performance comparing to the optimal alternative.
  \item We propose a scalable algorithm based on a competitive mechanism which enables a large number of UEs to access the network and select the appropriate APs for service.
  \item We propose two pilot assignment schemes for structured massive access, namely User-Group scheme and interference-based K-means (IB-KM) scheme.
        Both of them are designed to suppress the mutual interference from the pilot sharing among UEs by partitioning the UEs in a proper manner, and shown to outperform the benchmarks.
  \item We propose a scalable fractional power control policy where a suitable tradeoff between fairness and average SE can be found by adjusting a parameter.
  \item We derive two novel closed-form SE expressions with maximum ratio (MR) combining, whereof one is suitable for arbitrarily fixed pilot assignment schemes and the other is dedicated to the random pilot switching scheme.
\end{enumerate}}

\subsection{Paper Outline and Notations}
The remainder of this paper is organized as follows.
Section \ref{sec:system} introduces the system model for scalable cell-free massive MIMO.
The proposed P-LSFD strategy and its closed-form SE expression with MR combining are also provided in this section.
Section \ref{sec:ap_sele} proposes a scalable algorithm for massive UEs to accessing the network and selecting APs for service.
Another closed-form SE expression with MR combining and random pilot switching is provided in Section \ref{sec:pilo_assi}, and two novel pilot assignment schemes are proposed.
The performance of the proposed structured massive access framework is numerically evaluated in Section \ref{sec:numerical results}.
Finally, the major conclusions and implications are drawn in Section \ref{sec:conclusion}.

Boldface lowercase letters, $\bf x$, denote column vectors and boldface uppercase letters, $\bf X$, denote matrices.
${\bf X}_{ij}$ and ${{\bf{X}}_{ \cdot j}}$ denote the entry $\left({i,j}\right)$ and the $j$th column of matrix $\bf X$, respectively.
The superscripts $^{\rm T}$, $^{\rm *}$, and $^{\rm H}$ denote transpose, conjugate, and conjugate transpose, respectively.
The $n \times n$ identity matrix is ${\bf I }_n$.
We use $ \buildrel \Delta \over = $ for definitions and ${\rm {diag}}\left({{\bf A}_{1} ,\ldots, {\bf A}_{n}}\right)$ for a block-diagonal matrix with the square matrices ${{\bf A}_{1} ,\ldots, {\bf A}_{n}}$ on the diagonal.
The multi-variate circularly symmetric complex Gaussian distribution with correlation matrix $\bf R$ is denoted ${\cal N}_{\mathbb C}\left({{\bf 0},{\bf R}}\right)$.
{The expected value of $\bf x$ is denoted as ${\mathbb E}\left\{{\bf x}\right\}$.}
We denote by ${\left\| {\bf x} \right\|}_2$ the Euclidean norm of $\bf x$.
{We use $\left|{\cal A}\right|$ and $ {\cal A} \left(n\right)$ to denote the cardinality and the $n$th element of the set $\cal A$, respectively.}

\section{System Setup}\label{sec:system}

\begin{figure}[t]
\centering
\includegraphics[scale=0.74]{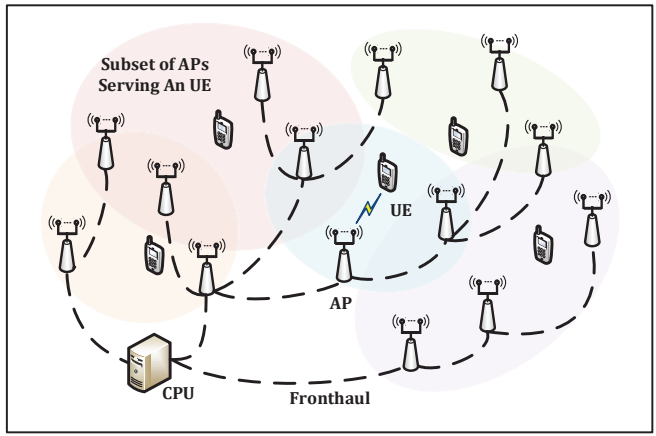}
\caption{A user-centric cell-free massive MIMO network, where each UE is served by as subset of APs.
\label{fig:system}}
\end{figure}

We consider a cell-free massive MIMO system consisting of $K$ single-antenna UEs and $L$ APs equipped with $N$ antennas.
As illustrated in Fig.~\ref{fig:system}, all APs are connected to a CPU in an arbitrary fashion. {We assume that the fronthaul connections are error-free since the focus of this paper is not on fronthaul provisioning.}
The channel between AP $l$ and UE $k$ is denoted as ${{\bf{h}}_{kl}} \in {{\mathbb{C}}^N}$.
The standard block fading model is considered \cite{bjornson2017massive}, where ${\bf{h}}_{kl}$ is constant in time-frequency blocks of $\tau_c$ channel uses.
In each block, an independent realization from a correlated Rayleigh fading distribution is drawn as ${\bf{h}}_{kl}  \sim {\cal{N}}_{\mathbb{C}}\left( {{\bf{0}},{\bf{R}}_{kl}}\right)$, where ${\bf{R}}_{kl}$ is the spatial correlation matrix describing the spatial property of the channel, and ${\beta _{kl}} \buildrel \Delta \over = {\rm{tr}}\left( {{{\bf{R}}_{kl}}} \right)/N$ is the large-scale fading coefficient that describes pathloss and shadowing.
The fading channels of different links are independently distributed. 
{We assume that deterministic information is known to the system; in particular, the spatial correlation matrices \{${\bf R}_{kl}$\} are available at the APs and the geographic locations of the APs is available at the CPU.}

{In order to achieve scalability in the system, we define a set of block-diagonal matrices ${\bf D}_{k}= {\rm{diag}}\left({{\bf D}_{k1} ,\ldots, {\bf D}_{kL}}\right) $, for $k = 1,\ldots,K $, where ${\bf D}_{kl} \in {\mathbb C}^{N \times N}$ is a diagonal matrix determining the antenna configuration at AP $l$ for UE $k$.}
More precisely, the $n$th diagonal entry of ${\bf D}_{kl}$ is $1$ if the $n$th antenna of AP $l$ is allowed to transmit to and decode signals from UE $k$ and $0$ otherwise.
{Moreover, we define a matrix ${\bf A} \in {\mathbb R}^{L \times K}$ specifying the \emph{AP selection} between UEs and APs, where the entry ${\bf A}_{kl}=1$ if ${\rm{tr}}\left( {{{\bf{D}}_{kl}}} \right) > 0$ and $0$ otherwise.
For the conciseness of mathematical descriptions, we denote by ${\cal M}_k = \left\{ {l: {\bf A}_{kl}=1 , l \in \left\{{ 1,\ldots,L }\right\}} \right\}$
the subset of APs serving UE $k$, and ${\cal D}_l = \left\{ {k: {\bf A}_{kl}=1, k \in \left\{{ 1,\ldots,K }\right\}} \right\}$ the subset of UEs served by AP $l$.}

For the uplink transmission, we have $\tau_p$ channel uses dedicated to pilots and the rest $\tau_c -\tau_p$ channel uses for payload data.
The two phases are described below.
Notice that the results of this paper are not limited in the systems operating in time-division duplex (TDD), but also apply to frequency-division duplex (FDD) mode, since the uplink works the same procedure in both duplex modes.

\subsection{Pilot Transmission and Channel Estimation}
We assume there are $\tau_p$ mutually orthogonal $\tau_p$-length pilot signals ${{\pmb{\phi }}_1}, \ldots ,{{\pmb{\phi }}_{\tau_p }}$ satisfying ${\left\| {{{\pmb{\phi }}_t}} \right\|^2} = \tau_p $, with $\tau_p$ being a constant independent of $K$.
{Every UE is assigned to a pilot when it accesses the network.
We consider a massive access scenario with a large number of UEs, in the sense that $K > \tau_p$. Hence, several UEs share the same pilot and these are referred to as \emph{pilot-sharing} UEs.
We denote by $t_k \in \left\{{ 1, \ldots , \tau_p }\right\}$ the index of the pilot assigned to UE $k$, and ${{\cal S}_k}$ the set of pilot-sharing UEs of UE $k$, including UE $k$ itself.
When the UEs in ${\cal{S}}_k$ transmit pilot ${{\pmb{\phi }}_{t_k}}$, AP $l$ receives the pilot signal ${\bf{y}}_{t_{k}l}^{\rm{p}} \in {{\mathbb C}^N}$ as \cite[Sec. 3]{bjornson2017massive}}
\begin{equation}\label{eq:received pilot signal in AP}
  {\bf{y}}_{{t_k}l}^{\rm{p}} = \sum\limits_{i \in {{\cal S}_k}} {\sqrt {{\tau _p}{p_i}} {{\bf{h}}_{il}}}  + {{\bf{n}}_{{t_k}l}},
\end{equation}
{where $p_i$ denotes the pilot transmit power of UE $i$ and ${{\bf{n}}_{{t_k}l}} \sim {{\cal N}_{\mathbb{C}}}\left( {{\bf{0}},{\sigma ^2}{{\bf{I}}_N}} \right)$ is the thermal noise.}
The MMSE estimate of ${\bf h}_{kl}$ for $k \in {\cal S}_k$ is given by \cite[Sec. 3]{bjornson2017massive}
\begin{equation}\label{eq:}
  {{{\bf{\hat h}}}_{kl}} = \sqrt {{\tau _p}{p_k}} {{\bf{R}}_{kl}}{\bf{\Psi }}_{{t_k}l}^{ - 1}{\bf{y}}_{{t_k}l}^{\rm{p}},
\end{equation}
where
\begin{equation}\label{eq:}
  {{\bf{\Psi }}_{{t_k}l}} = {\mathbb E}\left\{ {{\bf{y}}_{{t_k}l}^{\rm{p}}{{\left( {{\bf{y}}_{{t_k}l}^{\rm{p}}} \right)}^{\rm{H}}}} \right\} = \sum\limits_{i \in {{\cal S}_k}} {{\tau _p}{p_i}{{\bf{R}}_{il}}}  + {\sigma ^2}{{\bf{I}}_N}
\end{equation}
is the correlation matrix of \eqref{eq:received pilot signal in AP}. The estimate ${\hat{\bf h}}_{kl}$ and estimation error ${\tilde{\bf h}}_{kl} = {{\bf h}}_{kl} -{\hat{\bf h}}_{kl}$ are independent vectors distributed as ${\hat{\bf h}}_{kl} \sim {\cal N}_{\mathbb C}\left({ {\bf 0},{\bf B}_{kl} }\right)$ and ${\tilde{\bf h}}_{kl} \sim {\cal N}_{\mathbb C}\left({ {\bf 0},{\bf C}_{kl} }\right)$, where
\begin{equation}\label{eq:estimate variance}
  {\bf B}_{kl} = {\mathbb E}\left\{{ {\hat{\bf h}}_{kl}{\hat{\bf h}}^{\rm H}_{kl} }\right\} = {\tau _p}{p_k}{{\bf{R}}_{kl}}{\bf{\Psi }}_{{t_k}l}^{ - 1}{{\bf{R}}_{kl}},
\end{equation}
\begin{equation}\label{eq:}
  {\bf C}_{kl} = {\mathbb E}\left\{{ {\tilde{\bf h}}_{kl}{\tilde{\bf h}}^{\rm H}_{kl} }\right\} = {{\bf{R}}_{kl}} - {{\bf{B}}_{kl}}.
\end{equation}
{Note that \eqref{eq:received pilot signal in AP} indicates that sharing pilot ${\pmb \phi}_{t_k}$ among the UEs in ${\cal S}_k$ generates mutual interference, and consequently degrades the system performance, which is the so-called \emph{pilot contamination}.}

\subsection{Uplink Data Transmission}
{During the uplink data transmission, AP $l$ receives the signal ${\bf y}_{l} \in {\mathbb C}^{N}$ from all UEs,} as
\begin{equation}\label{eq:}
  {\bf{y}}_l^{} = \sum\limits_{i = 1}^K {{{\bf{h}}_{il}}{s_i}}  + {{\bf{n}}_l},
\end{equation}
where $s_i \sim {\cal N}_{\mathbb C}\left({ 0,p_i }\right)$ is the signal transmitted from UE $i$ with power $p_i$ and ${{\bf{n}}_l} \sim {{\cal N}_{\mathbb C}}\left( {{\bf{0}},{\sigma ^2}{{\bf{I}}_N}} \right)$ is the independent receiver noise.

{For the large-scale network deployment, we prefer to offload most of the computational tasks to the APs to avoid overloading the CPU.
More specifically, every AP preprocesses its signal by computing local estimates of the data and then passes them to the CPU for final decoding, which is the so-called \emph{LSFD}. Although all APs can physically receive the signal from all UEs, only the APs in the set ${\cal M}_k $ take part in the signal detection for UE $k$ due to the AP selection.
We denote by ${\bf a}_{kl} \in {\mathbb C}^{N}$ the combining vector selected by AP $l$ for UE $k$, where $k \in {\cal D}_l$.}
Then, the local estimate of $s_k$ is given by
\begin{equation}\label{eq:local estimate of s_k}
{{\tilde s}_{kl}} = {\bf{a}}_{kl}^{\rm{H}}{{\bf{D}}_{kl}}{\bf{y}}_l = {\bf{a}}_{kl}^{\rm{H}}{{\bf{D}}_{kl}}{{\bf{h}}_{kl}}{s_k} + {\bf{a}}_{kl}^{\rm{H}}{{\bf{D}}_{kl}}\sum\limits_{i = 1,\;i \ne k}^K {{{\bf{h}}_{il}}{s_i}}  + {\bf{a}}_{kl}^{\rm{H}}{{\bf{D}}_{kl}}{{\bf{n}}_l}.
\end{equation}
Any combining vector can be adopted in the above expression.
MR combining with ${\bf a}^{\rm {MR}}_{kl} = {\hat {\bf h}}_{kl}$ was considered in \cite{nayebi2016performance}, while \cite{bjornson2019scalable} has recently advocated for using
the local partial MMSE (LP-MMSE) combining
\begin{equation}\label{eq:}
    {\bf{a}}_{kl}^{{\rm{LP - MMSE}}} = {p_k}\left( {\sum\limits_{i \in {{\cal D}_l}} {{p_i}\left( {{{{\bf{\hat h}}}_{il}}{\bf{\hat h}}_{il}^{\rm{H}} + {{\bf{C}}_{il}}} \right)}  + {\sigma ^2}{{\bf{I}}_N}} \right)^{-1}{{{\bf{\hat h}}}_{kl}}.
\end{equation}

Then the local estimates $\left\{{ {{\tilde s}_{kl}} }\right\}$ are sent to the CPU, where they are linearly combined by using the weights $\left\{{ {{ w}_{kl}} }\right\}$ to obtain ${{\hat s}_k} = \sum\limits_{l = 1}^L {w_{kl}^*{{\tilde s}_{kl}}} $, which is eventually used to decode $s_k$.
From \eqref{eq:local estimate of s_k}, we have the final estimate of $s_k$, as
\begin{equation}\label{eq:estimate of s_k}
    {{\hat s}_k} = {\bf{a}}_k^{\rm{H}}{\bf{W}}_k^{\rm{H}}{{\bf{D}}_k}{{\bf{h}}_k}{s_k} + \sum\limits_{i = 1,\;i \ne k}^K {{\bf{a}}_k^{\rm{H}}{\bf{W}}_k^{\rm{H}}{{\bf{D}}_k}{{\bf{h}}_i}{s_i}}  + {\bf{a}}_k^{\rm{H}}{\bf{W}}_k^{\rm{H}}{{\bf{D}}_k}{\bf{n}},
\end{equation}
{where ${{\bf{W}}_k} = {\rm{diag}}\left( {{w_{k1}}{{\bf{I}}_N}, \ldots ,{w_{kL}}{{\bf{I}}_N}} \right) \in {{\mathbb C}^{\left( {L N} \right) \times \left( {L  N} \right)}}$.}

{Since the CPU does not have the knowledge of channel estimates}, we utilize the so-called \emph{use-and-then-forget} (UatF) bound \cite[Th. 4.4]{bjornson2017massive} to obtain the achievable SE.

\begin{lemm}\label{lemm:}
The achievable SE for UE $k$ of cell-free massive MIMO is
\begin{equation}\label{eq:}
{\rm{SE}}_k^{} = \left( {1 - \frac{{{\tau _p}}}{{{\tau _c}}}} \right){\log _2}\left( {1 + {\rm{SINR}}_k^{}} \right),
\end{equation}
where ${\rm{SINR}}_k$ is given by
\begin{align}\label{eq:UatF SINR_k}\notag
{\rm{SINR}}_k & = \frac{{{p_k}{{\left| {{\mathbb E}\left\{ {{\bf{a}}_k^{\rm{H}}{\bf{W}}_k^{\rm{H}}{{\bf{D}}_k}{{\bf{h}}_k}} \right\}} \right|}^2}}}{{\sum\limits_{i = 1}^K {{p_i}\underbrace {{\mathbb E}\left\{ {{{\left| {{\bf{a}}_k^{\rm{H}}{\bf{W}}_k^{\rm{H}}{{\bf{D}}_k}{{\bf{h}}_i}} \right|}^2}} \right\}}_{{\rm{\bf E}}_{ik}^{\left( 2 \right)}}}  - {p_k}\underbrace {{{\left| {{\mathbb E}\left\{ {{\bf{a}}_k^{\rm{H}}{\bf{W}}_k^{\rm{H}}{{\bf{D}}_k}{{\bf{h}}_k}} \right\}} \right|}^2}}_{{{\left| {{\rm{\bf E}}_k^{\left( 1 \right)}} \right|}^2}} + {\sigma ^2}\underbrace {{\mathbb E}\left\{ {{{\left\| {{{\bf{D}}_k}{\bf{W}}_k^{\rm{H}}{{\bf{a}}_k}} \right\|}^2}} \right\}}_{{\rm{\bf E}}_k^{\left( 3 \right)}}}}\\
 &= \frac{{{p_k}{{\left| {{\bf{w}}_k^{\rm{H}}{{\bf{v}}_k}} \right|}^2}}}{{{\bf{w}}_k^{\rm{H}}\left( {\sum\limits_{i = 1}^K {{p_i}{\bf{\Lambda }}_{ki}^{\left( 1 \right)}}  - {p_k}{{\bf{v}}_k}{\bf{v}}_k^{\rm{H}} + {\sigma ^2}{\bf{\Lambda }}_k^{\left( 2 \right)}} \right){{\bf{w}}_k}}},
\end{align}
where
\begin{equation}\label{eq:}
{{\bf{w}}_k} = {\left[ {{w_{kl}}, \ldots ,{w_{kL}}} \right]^{\rm{T}}},
\end{equation}
\begin{equation}\label{eq:}
{{\bf{v}}_k} = {\left[ {{\mathbb E}\left\{ {{\bf{a}}_{k1}^{\rm{H}}{{\bf{D}}_{k1}}{{\bf{h}}_{k1}}} \right\}, \ldots ,{\mathbb E}\left\{ {{\bf{a}}_{kL}^{\rm{H}}{{\bf{D}}_{kL}}{{\bf{h}}_{kL}}} \right\}} \right]^{\rm{T}}},
\end{equation}
\begin{equation}\label{eq:}
{\bf{\Lambda }}_{ki}^{\left( 1 \right)} = \left[ {\mathbb E}{\left\{ {{\bf{a}}_{kl}^{\rm{H}}{{\bf{D}}_{kl}}{{\bf{h}}_{il}}{\bf{h}}_{ij}^{\rm{H}}{{\bf{D}}_{kj}}{{\bf{a}}_{kj}}} \right\}:l,j= 1,\ldots,L} \right],
\end{equation}
\begin{equation}\label{eq:}
{\bf{\Lambda }}_k^{\left( 2 \right)} \!=\! {\rm{diag}}\left( {{\mathbb E}\left\{ {{{\left\| {{{\bf{D}}_{k1}}{{\bf{a}}_{k1}}} \right\|}^2}} \right\}, \ldots ,{\mathbb E}\left\{ {{{\left\| {{{\bf{D}}_{kL}}{{\bf{a}}_{kL}}} \right\|}^2}} \right\}} \right),
\end{equation}
and the expectations are with respect to all sources of randomness.
\end{lemm}
\begin{IEEEproof}
It follows the similar approach as in \cite[The. 4.4]{bjornson2017massive}, but for the received signal in \eqref{eq:estimate of s_k}.
\end{IEEEproof}

The structure of \eqref{eq:UatF SINR_k} is a generalized Rayleigh quotient with respect to ${\bf w}_{k}$.
{As a consequence}, the maximum value of $\rm{SINR}_k$ is achieved as \cite[Lem. B.10]{bjornson2017massive}
\begin{equation}\label{eq:}
{\rm{SINR}}_k^{} = {p_k}{\bf{v}}_k^{\rm{H}}{\left( {\sum\limits_{i = 1}^K {{p_i}{\bf{\Lambda }}_{ki}^{\left( 1 \right)}}  - {p_k}{{\bf{v}}_k}{\bf{v}}_k^{\rm{H}} + {\sigma ^2}{\bf{\Lambda }}_k^{\left( 2 \right)}} \right)^{ - 1}}{{\bf{v}}_k},
\end{equation}
with the optimal LSFD weight
\begin{equation}\label{eq:LSFD}
{{\bf{w}}^{\rm{LSFD}}_k} = {\left( {\sum\limits_{i = 1}^K {{p_i}{\bf{\Lambda }}_{ki}^{\left( 1 \right)}} + {\sigma ^2}{\bf{\Lambda }}_k^{\left( 2 \right)}} \right)^{ - 1}}{{\bf{v}}_k}.
\end{equation}

\renewcommand\arraystretch{2}
\begin{table*}[tp]
  \centering
  \fontsize{9}{9}\selectfont
  \caption{Fronthaul load related to the statistical parameters and the computational complexity of the weighting vector. }
  \label{tab:lsfd}
    \begin{tabular}{|p{1.5cm}<{\centering}|p{6.5cm}<{\centering}|p{6.5cm}<{\centering}|}
    \hline
        Scheme  &Fronthaul load (complex scalars)    &Computational complexity     \cr\hline
    \hline
        LSFD    & $K\left| {{{\cal M}_k}} \right|{\rm{ + }}\left( {{{\left| {{{\cal M}_k}} \right|}^2}{K^2}{\rm{ + }}K\left| {{{\cal M}_k}} \right|} \right)/2$            & $\left( {\frac{{{{\left| {{{\cal M}_k}} \right|}^2} + \left| {{{\cal M}_k}} \right|}}{2}} \right)K + \frac{{{{\left| {{{\cal M}_k}} \right|}^3} - \left| {{{\cal M}_k}} \right|}}{3} + {\left| {{{\cal M}_k}} \right|^2}$  \cr\hline

        P-LSFD  & $\left| {{{\cal P}_k}} \right|\left| {{{\cal M}_k}} \right|{\rm{ + }}\left( {{{\left| {{{\cal M}_k}} \right|}^2}{{\left| {{{\cal P}_k}} \right|}^2}{\rm{ + }}\left| {{{\cal P}_k}} \right|\left| {{{\cal M}_k}} \right|} \right)/2{\rm{ }}$           & $\left( {\frac{{{{\left| {{{\cal M}_k}} \right|}^2} + \left| {{{\cal M}_k}} \right|}}{2}} \right)\left| {{{\cal P}_k}} \right| + \frac{{{{\left| {{{\cal M}_k}} \right|}^3} - \left| {{{\cal M}_k}} \right|}}{3} + {\left| {{{\cal M}_k}} \right|^2}$  \cr\hline
    \end{tabular}
\end{table*}

{The fronthaul load required to gather all the statistical matrices for computing the LSFD vector in \eqref{eq:LSFD} and the related computational complexity are summarized in Table \ref{tab:lsfd}. Clearly,  they grow very fast with the size of the network, making the implementation of the optimal LSFD unscalable.}

{To achieve the implementation, we propose to use the alternative P-LSFD vector as}
\begin{equation}\label{eq:partial LSFD}
{\bf{w}}_k^{{\rm{P - LSFD}}} = {\left( {\sum\limits_{i \in {{\cal P}_k}} {{p_i}{\bf{\Lambda }}_{ki}^{\left( 1 \right)}}  + {\sigma ^2}{\bf{\Lambda }}_k^{\left( 2 \right)}} \right)^{ - 1}}{{\bf{v}}_k},
\end{equation}
{where ${{\cal P}_k} = \left\{ {i:{{\bf{A}}_{kl}}{{\bf{A}}_{il}} \ne 0,l \in \left\{ {1, \ldots ,L} \right\}} \right\}$
is the index set of the UEs which are served by partially the same APs as UE $k$. Only those UEs in ${{\cal P}_k}$ might cause substantial interference to UE $k$.}
Note that $\left|{ {\cal P}_{k} }\right|  \le  \left({ {\tau_p - 1 }}\right)\left|{ {\cal M}_{k} }\right|+1$, where the upper bound is achieved
in the unlikely case that all the APs in ${\cal M}_{k}$ serve UE $k$ but otherwise serve entirely different sets of UEs.
Importantly, the upper bound is independent of $K$.
The fronthaul load related to the statistical parameters and the total number of complex multiplications required by P-LSFD is given in Table \ref{tab:lsfd}. {It is important to note that} the proposed P-LSFD is a scalable strategy whose complexity does not grow with $K$.

{The expectations in \eqref{eq:UatF SINR_k} cannot be computed in closed-form when using LP-MMSE, but can be easily computed using Monte-Carlo simulations.
Similar to \cite[Cor. 4.5]{bjornson2017massive}, we can obtain the following closed-form expression as a simple baseline when using MR combining.}

\begin{lemm}\label{lemm:MR}
If MR combining with ${\bf a}^{\rm {MR}}_{kl} = {\hat {\bf h}}_{kl}$ is used, the expectations in \eqref{eq:UatF SINR_k} become
\begin{equation}\label{eq:}
 {{\rm{\bf E}}_k^{\left( 1 \right)}} = {\bf{w}}_k^{\rm{H}}{{\bf{u}}_{kk}},
\end{equation}
\begin{equation}\label{eq:}
{{\rm{\bf E}}_k^{\left( 3 \right)}} = {\bf{w}}_k^{\rm{H}}{\bf{\Omega }}_k^{\left( 2 \right)}{{\bf{w}}_k},
\end{equation}
and
\begin{equation}\label{eq:}
{{\rm{\bf E}}_{ik}^{\left( 2 \right)}} = {\bf{w}}_k^{\rm{H}}{\bf{\Omega }}_{ki}^{\left( 1 \right)}{{\bf{w}}_k} +
 {\begin{cases}
{\frac{{{p_i}}}{{{p_k}}}{\bf{w}}_k^{\rm{H}}{{\bf{u}}_{ki}}{\bf{u}}_{ki}^{\rm{H}}{{\bf{w}}_k}}&{{\rm{if}}\;i \in {{\cal S}_k}},\\
0&{{\rm{otherwise}}},
\end{cases}}
\end{equation}
where
\begin{equation}\label{eq:}
{{\bf{u}}_{ki}} = {\left[ {{\rm{tr}}\left( {{{\bf{D}}_{k1}}{{\bf{B}}_{k1}}{\bf{R}}_{k1}^{ - 1}{{\bf{R}}_{i1}}} \right), \cdots ,{\rm{tr}}\left( {{{\bf{D}}_{kL}}{{\bf{B}}_{kL}}{\bf{R}}_{kL}^{ - 1}{{\bf{R}}_{iL}}} \right)} \right]^{\rm{T}}},
\end{equation}
\begin{equation}\label{eq:}
{\bf{\Omega }}_{ki}^{\left( 1 \right)} = {\rm{diag}}\left\{ {{\rm{tr}}\left( {{{\bf{D}}_{k1}}{{\bf{B}}_{k1}}{{\bf{R}}_{i1}}} \right), \ldots ,{\rm{tr}}\left( {{{\bf{D}}_{kL}}{{\bf{B}}_{kL}}{{\bf{R}}_{iL}}} \right)} \right\},
\end{equation}
and
\begin{equation}\label{eq:}
{\bf{\Omega }}_k^{\left( 2 \right)} = {\rm{diag}}\left\{ {{\rm{tr}}\left( {{{\bf{D}}_{kl}}{{\bf{B}}_{kl}}} \right), \cdots ,{\rm{tr}}\left( {{{\bf{D}}_{kl}}{{\bf{B}}_{kl}}} \right)} \right\}.
\end{equation}
\end{lemm}
\begin{IEEEproof}
It follows the similar approach as in \cite[Cor. 4.5]{bjornson2017massive}, but for the received signal in \eqref{eq:estimate of s_k}.
\end{IEEEproof}

\section{Initial Access and AP Selection}\label{sec:ap_sele}
{When UE $k$ accesses the network, it selects its serving APs, i.e., the APs in ${\cal M}_k$.} However, it cannot make this choice entirely freely since each AP only supports a limited number of UEs \cite{bjornson2019scalable}. {More precisely, each AP can only manage $\tau_p$ UEs, to avoid strong pilot contamination.
Therefore, we adopt the following key assumption from \cite{bjornson2019scalable}.}
\begin{assu}\label{assu:1}
Each AP serves at most one UE per pilot and uses all its $N$ antennas to serve these UEs.
\end{assu}
The above assumption implies that $\left|{ {\cal D}_l }\right| \le \tau_p$ and
\begin{equation}\label{eq:}
  {\bf D}_{kl} = \begin{cases}
  {\bf I}_{N}&{\rm{if}}\ k \in {\cal D}_l\\
  {\bf 0}_{N}&\rm{otherwise}
  \end{cases},
\end{equation}
for ${l = 1, \ldots ,L}$.

{In order to satisfy Assumption \ref{assu:1} and guarantee every UE at least has one serving AP, we develop an algorithm based on a competitive mechanism.
The main idea is that UE $k$ needs to compete for AP $l$ with $\tau_p$ UEs that might already be served by AP $l$.}
We denote by $k^*$ the index of the UE with the smallest large-scale fading coefficient in ${ \left\{{k}\right\} \cup {\cal D}_l }$.
{UE $k$ succeeds if $k \ne k^*$.
Then UE $k^*$ puts $l$ into its \emph{blacklist} ${\cal B}_{k^*}\subset \left\{{1,\ldots,L }\right\}$, which means AP $l$ is no longer available for UE $k^*$.
This is reasonable since the UEs that have won the competition have better channel conditions than UE $k^*$, and thus UE $k^*$ cannot win any competition regarding AP $l$.
Moreover, if $\left|{\cal B}_{k^*}\right|$ reaches $L-1$, which means UE ${k^*}$ has lost every competition it participated in, then UE ${k^*}$ is added into the list ${{\cal L}_{\overline {{\rm{UE}}} }}$ and assigned to the only AP that is left; consequently, UE ${k^*}$ no longer needs to participate in another competition.
${{\cal L}_{\overline {{\rm{UE}}} }}$ prevents the UEs in weak channel conditions from being abandoned.
We denote by ${\cal L}_{\rm{UE}}$ the list in ascending order, which comprises the indices of the UEs which have not finished their AP selections yet.}
The algorithm initiates with ${\cal L}_{\rm{UE}} = \left\{1,\ldots,K \right\}$, ${\cal L}_{\overline {\rm{UE}}} = \emptyset$, $\left\{ {\cal M}_k = \emptyset: k=1,\ldots,K \right\}$, and $\left\{ {\cal B}_k = \emptyset: k=1,\ldots,K \right\}$.

Our proposed AP selection algorithm operates through the following steps.
\begin{enumerate}
  \item
    {UE $k= {\cal L}_{\rm{UE}} \left( {1} \right)$ measures its large-scale fading coefficients with the APs in ${\cal L}_{{\rm{AP}},{k}}$, where ${\cal L}_{{\rm{AP}},{k}} = \left\{1,\ldots,L \right\}/\left\{{ {{{\cal M}_{k}}} \cup {{{\cal B}_{k}}}}\right\}$ is the list comprising the indices of the APs which are available for UE $k$.}

  \item
    {UE $k$ finds the AP
    \begin{equation}\label{eq:bestAP}
      l = \arg {\max _{j \in {{\cal L}_{{\rm{AP}},{k}}}}}{\beta _{kj}}
    \end{equation}
    If $\left|{\cal D}_l\right|< \tau_p $, UE $k$ takes AP $l$ as its serving AP by ${\cal M}_{k} \cup \left\{{l}\right\}$, and repeats Step 2) to seek for more APs; otherwise, a competition is needed, which is elaborated in Step 3).}

  \item
      A competition occurs when UE $k$ attempts to select AP $l$ while AP $l$ already has $\tau_p$ UEs in ${\cal D}_l$.
      {The principle is that AP $l$ gives priority to the UEs in stronger channel conditions.}
      Therefore, AP $l$ finds the ``weakest" UE
      \begin{equation}\label{eq:}
      k^* = \arg {\min _{i \in { \left\{{k}\right\} \cup {\cal D}_l / {\cal L}_{\overline {\rm{UE}}} }}}{\beta _{il}}.
      \end{equation}

      {If $k^* = k$, UE $k$ puts $l$ into ${\cal B}_k$;
      otherwise, UE $k$ succeeds UE $k^*$ in ${\cal D}_l$, and UE $k^*$ puts $l$ into ${\cal B}_{k^*}$.
      After the competition, UE $k$ goes back to Step 2) for another available AP, until ${\cal L}_{{\rm{AP}},k} = \emptyset$ or $k \in {\cal L}_{\overline {\rm{UE}}}$.
      In the case of $k \in {\cal L}_{\overline {\rm{UE}}}$, UE $k$ selects whatever AP left in ${\cal L}_{{\rm{AP}},k}$.
      If the only AP $l'$ left in ${\cal L}_{{\rm{AP}},k}$ already has $\tau_p$ UEs in ${\cal D}_{l'}$, then AP $l'$ turns to serve UE $k$ instead of UE}
      \begin{equation}\label{eq:}
      k' = \arg {\min _{i \in {{\cal D}_{{{l'}}}}/{{\cal L}_{\overline {{\rm{UE}}} }}}}{\beta _{il}}.
      \end{equation}
      By then, UE $k$ finishes its AP selection and is moved from ${\cal L}_{\rm{UE}}$ by ${\cal L}_{\rm{UE}}/\left\{k \right\}$.
  \item
    Go back to Step 1) for the next UE, until ${\cal L}_{\rm{UE}} = \emptyset$.
\end{enumerate}

{Based on the results of the AP selection, we construct the matrix ${\bf A}$.}
The pseudo code of this algorithm is given in Algorithm \ref{algo:ap selection}.
\begin{algorithm}[h]
\label{algo:ap selection}
\caption{Initial Access and AP Selection.}

\KwIn{$\left\{ {\beta}_{kl} \right\}$, ${\cal L}_{\rm{UE}}$, ${{\cal L}_{\overline {{\rm{UE}}} }}$, $\left\{ {\cal M}_{k} \right\}$, $\left\{ {\cal B}_{k} \right\}$, $\left\{ {\cal D}_{l} \right\}$}

\KwOut{$\left\{ {\cal M}_{k} \right\}$}

\For{$k \in {\cal L}_{\rm{UE}}$}
{
    \Repeat(){{\bf{break}}}
    {
        {${\cal L}_{{\rm{AP}},k} \leftarrow \left\{1,\ldots,L \right\}/\left\{{ {{{\cal M}_{k}}} \cup {{{\cal B}_{k}}}}\right\}$};\\
        \If{{${\cal L}_{{\rm{AP}},k} = \emptyset $}}
        {
            {\bf{break}};\\
        }
        \Else
        {
            \If{$k \in {{\cal L}_{\overline {{\rm{UE}}} }} $}
            {
                {${\cal M}_{k} \leftarrow \left\{l'\right\} = {\cal L}_{{\rm{AP}},k}$;}\\
                \If{$\left|{\cal D}_{l'}\right| = \tau_p$}
                {
                    {$k' = \arg {\min _{i \in {\cal D}_{l'}/{\cal L}_{\overline {\rm{UE}} }}}{\beta _{il}}$};\\
                    {${\cal M}_{k'} \leftarrow {\cal M}_{k'} / \left\{{l'}\right\}$};\\
                }
                {\bf{break}};\\
            }
            \Else
            {
                {$l = \arg {\max _{j \in {{\cal L}_{{\rm{AP}},k}}}}{\beta _{kj}}$};\\
                ${\cal M}_{k} \leftarrow {\cal M}_{k} \cup \left\{{l}\right\}$;\\
                \If{$\left|{\cal D}_l\right| > \tau_p$}
                {
                    $k^* = \arg {\min _{i \in {{\cal D}_l/{\cal L}_{\overline {\rm{UE}} }}}}{\beta _{il}}$;\\
                    ${\cal B}_{k^*} \leftarrow {\cal B}_{k^*} \cup \left\{{l}\right\}$;\\
                    \If{$\left|{{\cal B}_{k^*}}\right| = L-1$}
                    {
                        ${{\cal L}_{\overline {{\rm{UE}}} }} \leftarrow {{\cal L}_{\overline {{\rm{UE}}} }} \cup \left\{{k^*}\right\}$;\\
                    }
                    ${\cal M}_{k^*} \leftarrow {\cal M}_{k^*} / \left\{{l}\right\}$;\\
                }
            }
        }
    }
}
{\bf{final}};\\
\end{algorithm}

\section{Pilot Assignment}\label{sec:pilo_assi}
{A proper pilot assignment improves the system performance by suppressing the pilot contamination, particularly, in the massive access scenario.
In this section, we derive a novel closed-form SE expression when random pilot switching is applied.
Meanwhile, we elaborate the drawback of random pilot assignment, and propose two novel pilot assignment schemes dedicating to suppressing the pilot contamination.}
\subsection{Random Pilot Assignment and Random Pilot Switching}\label{subsec:}
When the random pilot assignment scheme is applied, every UE in the network is assigned a pilot at random from $\tau_p$ orthogonal pilots and uses it in all blocks.
Random pilot switching is another approach to assign pilots, in which each UE does not pick one pilot at random, but switches between pilots in a random fashion over blocks to average over the pilot contamination \cite{bjornson2016deploying}.
When random pilot switching is applied, the pilot-sharing UEs for UE $k$ will vary.
We use a binary random variable
\begin{equation}\label{eq:}
{\chi _{ik}} =  {\begin{cases}
1&{{\rm{if}}\;i \in {{\cal S}_k}},\\
0&{{\rm{otherwise}}},
\end{cases}} \quad i = 1, \ldots ,K
\end{equation}
instead of ${\cal P}_k$ to indicate whether a UE $i$ is a pilot-sharing UE of UE $k$ or not, since it is easier to define the statistics of ${\chi _{ik}}$ than ${\cal P}_k$.
The probability of ${\chi _{ik}} = 1 $ is $\frac{1}{{{\tau _p}}}$ and $1 - \frac{1}{{{\tau _p}}}$ otherwise.
With this notation, the despreaded pilot signal received at AP $l$ in \eqref{eq:received pilot signal in AP} can be rewritten as
\begin{equation}\label{eq:}
{\bf{y}}_{{t_k}l}^{\rm{p}} = \sqrt {{\tau _p}{p_k}} {{\bf{h}}_{kl}} + \sum\limits_{i = 1,\;i \ne k}^K {{\chi _{ik}}\sqrt {{\tau _p}{p_i}} {{\bf{h}}_{il}}}  + {{\bf{n}}_{{t_k}l}}.
\end{equation}
As a consequence, we have
\begin{equation}\label{eq:pilot signal variance random}
{{\bf{\Psi }}_{{t_k}l}} ={\mathbb E}_{\left\{{\bf h}\right\}}\left\{ {{\bf{y}}_{{t_k}l}^{\rm{p}}{{\left( {{\bf{y}}_{{t_k}l}^{\rm{p}}} \right)}^{\rm{H}}}} \right\}= {\tau _p}{p_k}{{\bf{R}}_{kl}} + \sum\limits_{i = 1,\;i \ne k}^K {{\chi _{ik}}{\tau _p}{p_i}{{\bf{R}}_{il}}}  + {\sigma ^2}{{\bf{I}}_N},
\end{equation}
where ${\mathbb E}_{\left\{{\bf h}\right\}}\left\{ \cdot \right\}$ denotes the expectation with respect to the channel and noise realizations.
Since no randomness appears in $\chi_{kk} = 1$, we rewrite the SINR expression in \eqref{eq:UatF SINR_k} as
\begin{equation}\label{eq:SINR_random}
{\rm{SINR}}_k = \frac{{{p_k}{{\left| {{\rm{\bf E}}_k^{\left( 1 \right)}} \right|}^2}}}{{\sum\limits_{i = 1,i \ne k}^K {{p_i}{\rm{\bf E}}_{ki}^{\left( 2 \right)}}  + {p_k}{\rm{\bf E}}_{kk}^{\left( 2 \right)} - {p_k}{{\left| {{\rm{\bf E}}_k^{\left( 1 \right)}} \right|}^2} + {\sigma ^2}{\rm{\bf E}}_k^{\left( 3 \right)}}}
\end{equation}
for the following derivation. A closed-form expression of SINR when using MR combining and random pilot switching is obtained as follows.
\begin{coro}\label{coro:random SINR}
If MR combining with ${\bf a}^{\rm {MR}}_{kl} = {\bf{B}}_{kl}^{{\rm{ - }}1}{{{\bf{\hat h}}}_{kl}}$ is used, the expectations in \eqref{eq:SINR_random} become
\begin{equation}\label{eq:expected signalx}
{{\rm{\bf E}}_k^{\left( 1 \right)}} = {\bf{w}}_k^{\rm{H}}{{{\bf{\bar u}}}_{kk}},
\end{equation}
\begin{equation}\label{eq:scalingx}
{\rm{\bf E}}_k^{\left( 3 \right)} = {\bf{w}}_k^{\rm{H}}{\bf{\bar \Omega }}_k^{\left( 2 \right)}{{\bf{w}}_k},
\end{equation}
\begin{equation}\label{eq:interference_kx}
{\rm{\bf E}}_{kk}^{\left( 2 \right)} = {\bf{w}}_k^{\rm{H}}\left( {{\bf{\bar \Omega }}_{kk}^{\left( 1 \right)} + {{{\bf{\bar u}}}_{kk}}{\bf{\bar u}}_{kk}^{\rm{H}}} \right){{\bf{w}}_k},
\end{equation}
and
\begin{equation}\label{eq:interferencex}
{\rm{\bf E}}_{ik}^{\left( 2 \right)} = {\bf{w}}_k^{\rm{H}}\left( {{\bf{\bar \Omega }}_{ki}^{\left( 1 \right)} + \frac{{{p_i}}}{{{\tau _p}{p_k}}}{{{\bf{\bar u}}}_{ki}}{\bf{\bar u}}_{ki}^{\rm{H}}} \right){{\bf{w}}_k}, i\ne k,
\end{equation}
where
\begin{equation}\label{eq:}
{{{\bf{\bar u}}}_{ki}} = {\left[ {{\rm{tr}}\left( {{{\bf{D}}_{k1}}{{\bf{R}}_{i1}}{\bf{R}}_{k1}^{ - 1}} \right), \ldots ,{\rm{tr}}\left( {{{\bf{D}}_{kL}}{{\bf{R}}_{iL}}{\bf{R}}_{kL}^{ - 1}} \right)} \right]^{\rm{T}}},
\end{equation}
\begin{equation}\label{eq:}
{\bf{\bar \Omega }}_{ki}^{\left( 1 \right)} = {\rm{diag}}\left\{ {{\rm{tr}}\left( {{{\bf{D}}_{kl}}{\bf{\bar B}}_{kl}^{ - 1}{{\bf{R}}_{il}}} \right), \ldots ,{\rm{tr}}\left( {{{\bf{D}}_{kl}}{\bf{\bar B}}_{kl}^{ - 1}{{\bf{R}}_{il}}} \right)} \right\},
\end{equation}
\begin{equation}\label{eq:}
{\bf{\bar \Omega }}_k^{\left( 2 \right)} = {\rm{diag}}\left\{ {{\rm{tr}}\left( {{{\bf{D}}_{k1}}{\bf{\bar B}}_{k1}^{ - 1}} \right), \ldots ,{\rm{tr}}\left( {{{\bf{D}}_{kL}}{\bf{\bar B}}_{kL}^{ - 1}} \right)} \right\},
\end{equation}
and
\begin{align}\label{eq:estimate variance bar}\notag
{\bf{\bar B}}_{kl}^{ - 1} &= {{\mathbb E}_{\left\{ \chi  \right\}}}\left\{ {{\bf{B}}_{kl}^{ - 1}} \right\}= \frac{1}{{{\tau _p}{p_k}}}{\bf{R}}_{kl}^{ - 1}\left( {\sum\limits_{i = 1,\;i \ne k}^K {{{\mathbb E}_{\left\{ \chi  \right\}}}\left\{ {{\chi _{ik}}} \right\}{\tau _p}{p_i}{{\bf{R}}_{il}}}  + {\tau _p}{p_k}{{\bf{R}}_{kl}} + {\sigma ^2}{{\bf{I}}_N}} \right){\bf{R}}_{kl}^{ - 1}\\
 &= \frac{1}{{{\tau _p}{p_k}}}{\bf{R}}_{kl}^{ - 1}\left( {{\tau _p}{p_k}{{\bf{R}}_{kl}} + \sum\limits_{i = 1,\;i \ne k}^K {{p_i}{{\bf{R}}_{il}}}  + {\sigma ^2}{{\bf{I}}_N}} \right){\bf{R}}_{kl}^{ - 1},
\end{align}
{with the fact that ${\mathbb E}\left\{ {{\chi _{ik}}} \right\} = \frac{1}{{{\tau _p}}}$, $i \ne k$, where ${\mathbb E}_{\left\{{{\chi }}\right\}}\left\{ \cdot \right\}$ denotes the expectation with respect to ${{\chi }}$.}
\end{coro}
\begin{IEEEproof}
 The proof follows the similar approach as in \cite[Appe. D]{ozdogan2019performance}, but the derivation is performed by first computing the expectations with respect to $\bf h$, then computing the expectations with respect to $\chi$.
\end{IEEEproof}
Note that the normalization of ${{{\bf{\hat h}}}_{kl}}$ with ${\bf{B}}_{kl}^{{\rm{ - }}1}$ in Corollary \ref{coro:random SINR} makes the expected channel gain equal to ${\bf{w}}_k^{\rm{H}}{{{\bf{\bar u}}}_{kk}}$ as in \eqref{eq:expected signalx}, and thereby enables us to derive the closed-form expressions in Corollary~\ref{coro:random SINR}.

We treat the closed-form SE expression obtained in Corollary \ref{coro:random SINR} as a ``worst'' case, since all UEs in the network are possibly suffering from strong pilot contamination in random pilot switch.
Therefore, we mainly consider the random pilot assignment, which is widely considered in previous works, as a benchmark. The reason is that two UEs that are close to each other will occasionally share the same pilot and then create strong mutual interference. This can be avoided by a structured pilot assignment.

\subsection{Interference-Based K-Means Pilot Assignment Scheme}\label{subsec:IB-KM}

A K-means-type pilot assignment scheme was proposed in \cite{attarifar2018random} and we call it geography-based K-means (GB-KM) pilot assignment since the geographic location of the UEs is utilized. Inspired by this scheme, we propose another K-means-type pilot assignment scheme where instead the distances between all UEs and APs are considered.
Note that no extra processing is needed for this distance information since it is intermediate when the APs and CPU obtain $\left\{ {\bf R}_{kl} \right\}$.
{Since this scheme aims to suppress the interference generated by the pilot-sharing UEs, we refer it to as IB-KM pilot assignment scheme.}
Before we elaborate the scheme, we first make the following key assumption.
\begin{assu}\label{assu:2}
The level to the inter-user interference generated by UE $i$ and UE $k$ is indicated by
\begin{equation}\label{eq:Dis}
 {\rm{Dis}}_{ik} = \left\| {{\rm{diag}}\left({{\bf{d}}_i}\right)  {{\bf A}_{\cdot i}} - {\rm{diag}}\left({{\bf{d}}_k}\right)  {{\bf A}_{\cdot k}}} \right\|_2^2,
\end{equation}
where ${{\bf{d}}_i} = {\left[ {{d_{i1}}, \ldots ,{d_{iL}}} \right]^{\rm{T}}}$ and ${{\bf A}_{\cdot i}}$ denote the distance and serving relationship between UE $i$ and all APs.
The smaller values of ${\rm{Dis}}_{ik}$ indicate the stronger inter-user interference could be generated if UE $i$ and UE $k$ share the same pilot.
\end{assu}
\begin{figure}[t]
\centering
\includegraphics[scale=0.6]{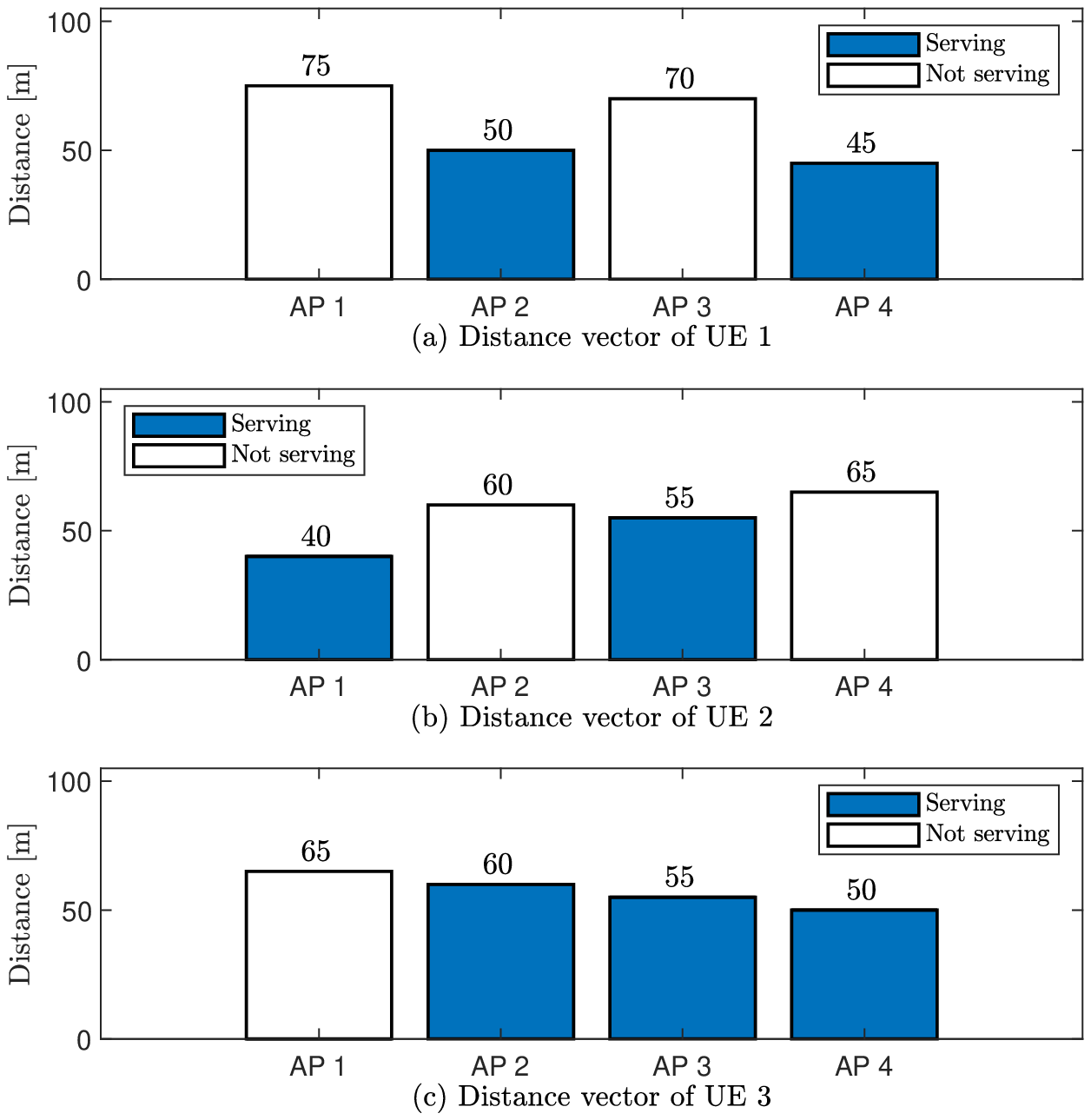}
\caption{An example of Assumption \ref{assu:2}. (a) UE $1$: ${\bf{d}}_1 = \left[75,50,70,45 \right]^{\rm T}$, ${\bf{A}}_{\cdot 1} = \left[0,1,0,1 \right]^{\rm T}$; (b) UE $2$: ${\bf{d}}_2 = \left[45,60,55,65 \right]^{\rm T}$, ${\bf{A}}_{\cdot 2} = \left[1,0,1,0 \right]^{\rm T}$; (c) UE $3$: ${\bf{d}}_3 = \left[65,60,55,50 \right]^{\rm T}$, ${\bf{A}}_{\cdot 3} = \left[0,1,1,1 \right]^{\rm T}$.
\label{fig:assu2}}
\end{figure}
{The rationale behind Assumption 2 is that the inter-user interference occurs when the pilot-sharing UEs communicate with the same AP. The strength of the interference depends on the signal power of the pilot-sharing UEs, which is mainly determined by the distances between the pilot-sharing UEs and the same AP when the channel distribution and the transmit power are roughly the same.}
A simple example with $3$ UEs and $4$ APs is provided in Fig.~\ref{fig:assu2} to explain Assumption \ref{assu:2}.
The distances between UE $k$ and its serving APs (i.e., ${\rm{diag}}\left({{\bf{d}}_k}\right)  {{\bf A}_{\cdot k}}$) are marked with ``Serving" in Fig.~\ref{fig:assu2}.
In the example, we can see that UE $2$ and UE $3$ are located in the similar positions but served by different subsets of APs (${\cal M}_2 = \left\{1,3 \right\}$ and ${\cal M}_3 = \left\{2,3,4 \right\}$).
When comparing the cases of these $3$ UEs in Fig.~\ref{fig:assu2}, we can conclude that UE $1$ and UE $2$ will generate less inter-user interference if they share the same pilot than UE $1$ and UE $3$.
{The reason is that UE $1$ and UE $2$ are served by disjoint subsets of APs while UE $1$ and UE $3$ have a common serving AP, i.e., AP $3$.}
Then we back to \eqref{eq:Dis} and find that ${\rm{Dis}}_{12} = 9150 > {\rm{Dis}}_{13} = 315$, since ${\rm{Dis}}_{ik}$ indicates the difference between service quality of UE $i$ and UE $k$ from the APs of their corresponding ${\cal M}_i$ and ${\cal M}_k$.

\begin{figure}[t]
\centering
\includegraphics[scale=1]{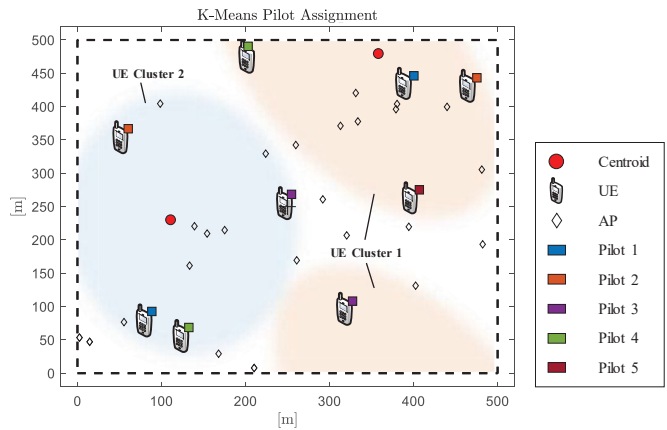}
\caption{A cell-free massive MIMO network with K-means pilot assignment, where $9$ UEs are separated into $2$ clusters center on $2$ centroids. $5$ pilots are reused in each clusters.
\label{fig:K-means}}
\end{figure}

Based on Assumption \ref{assu:2}, the basic idea of the IB-KM pilot assignment scheme is that the $K$ UEs are separated into $\left\lceil {K/{\tau _p}} \right\rceil $ disjoint clusters centering on $\left\lceil {K/{\tau _p}} \right\rceil $ \emph{centroids}, whose minimum ${\rm{Dis}}$ with each other is as large as possible.
{When the APs are deployed, the location of these centroids can be trained with a large number of \emph{points} randomly locating in the coverage area ,which could be generated by the MATLAB function ``{\tt{rand}}" \cite{attarifar2018random}.}
Every such cluster comprises at most $\tau_p$ UEs, which have the smallest values of the ${\rm{Dis}}$ with the corresponding centroid.
UEs in the same cluster are assigned mutually orthogonal pilots, as shown in Fig.~\ref{fig:K-means}.
{The algorithm initiates with $\left\{ {\cal C}_m = \emptyset: m=1,\ldots,\left\lceil {K/{\tau _p}} \right\rceil \right\}$ and $\varepsilon = 0.001$. Note that the distance between APs and UEs $\left\{ {{{\bf{d}}_k}} \right\}$ is generated when the spatial correlation matrices $\{{\bf R}_{kl}\}$ are generated, which depends on the simulation setup.}

Our proposed IB-KM pilot assignment scheme operates through the following steps.
\begin{enumerate}
  \item Arbitrarily generate $K_{\rm p}$ points and $\left\lceil {K/{\tau _p}} \right\rceil$ centroids in the coverage area, where $K_{\rm p}$ is a large number.
      Each point and centroid measures its distance with all APs, generates distance vector ${{\bf{ d}}'_p} = {\left[ {{{ d}'_{p1}}, \ldots ,{{ d}'_{pL}}} \right]^{\rm{T}}}$, $p = 1, \ldots ,{K_{\rm{p}}}$ and ${{\pmb \mu} _m} = {\left[ {{\mu _{m1}}, \ldots ,{\mu _{mL}}} \right]^{\rm{T}}}$, $m = 1, \ldots ,\left\lceil {K/{\tau _p}} \right\rceil $, respectively.
  \item Each point selects the centroid
      \begin{equation}\label{eq:}
       m^* = \arg {\min _{1\le m \le \left\lceil {K/{\tau _p}} \right\rceil}}{\left\| {{{\bf{ d}}'_p} - {{\pmb{\mu }}_m}} \right\|^2_2}, \ p = 1,\ldots,K_{\rm p},
      \end{equation}
      and join the corresponding cluster ${\cal C}_{m^*}$.
  \item Each centroid updates its distance vector as
       \begin{equation}\label{eq:}
       {{\pmb{\mu }}'_m} = \frac{1}{{\left| {{{\cal C}_m}} \right|}}\sum\limits_{{p} \in {{\cal C}_m}} {{{\bf{ d}}'_p}},\  m=1,\ldots,\left\lceil {K/{\tau _p}} \right\rceil,
       \end{equation}
       and go back to step 1), until
       \begin{equation}\label{eq:}
       {\max _{1\le m \le \left\lceil {K/{\tau _p}} \right\rceil}}{\left\| {{{\pmb{\mu }}'_{m}} - {{\pmb{\mu }}_m}} \right\|^2_2}< \varepsilon,
       \end{equation}
       where $\varepsilon$ is a small number.
  \item Each UE generates its distance vector ${{\bf{d}}_i} = {\left[ {{d_{i1}}, \ldots ,{d_{iL}}} \right]^{\rm{T}}}$, $i = 1, \ldots ,{K}$.
  \item Each UE selects the centroid
      \begin{equation}\label{eq:find centroid}
       m^* = \arg {\min _{1\le m \le \left\lceil {K/{\tau _p}} \right\rceil}}{\left\| {{\rm{diag}}\left({{\bf{d}}_i}\right)  {{\bf A}_{\cdot i}} - {{\pmb{\mu }}_m}} \right\|^2_2}, \ i = 1,\ldots,K,
      \end{equation}
      and join in the corresponding cluster ${\cal C}_{m^*}$.
      A competition mechanism similar to the one in Algorithm \ref{algo:ap selection} could be applied if a generic centroid $m$ is selected by more than $\tau_p$ UEs.
      Or more succinctly, each cluster chooses $\tau_p$ UEs with the smallest values of $\rm{Dis}$ with the corresponding centroid in sequence, until all UEs are allocated into $\left\lceil {K/{\tau _p}} \right\rceil $ disjoint clusters; a UE only can be chosen by one cluster.
  \item Find a cluster with $\tau_p$ UEs and arbitrarily assign the UEs $\tau_p$ mutually orthogonal pilots.
        Without loss of generality, we assume $\left|{\cal C}_1\right| = \tau_p$ and assign the UEs in ${\cal C}_1$ pilots $\left\{ {{\pmb{\phi }}_1},\ldots,{{\pmb{\phi }}_{\tau_p}}  \right\}$.
  \item Each UE in ${\cal C}_1$ finds UE
      \begin{equation}\label{eq:find UE}
       i^* = \arg {\max _{i \in {\cal C}_m}}{\left\| {{\rm{diag}}\left({{\bf{d}}_i}\right)  {{\bf A}_{\cdot i}} - {\rm{diag}}\left({{\bf{d}}_k}\right)  {{\bf A}_{\cdot k}}} \right\|^2_2},\  k \in {\cal C}_1,
      \end{equation}
      in ${\cal C}_m$, $m = 2,\ldots,\left\lceil {K/{\tau _p}} \right\rceil$, and shares pilot with this UE.
      If a UE $i^*$ in ${\cal C}_m$ is selected by multiple UEs in ${\cal C}_1$, then only the UE, whose value of $\rm{Dis}$ with UE $i^*$ is the largest, shares pilot with UE $i^*$; the rest UEs find another UE based on \eqref{eq:find UE}, until each UE in the network are assigned a pilot.
\end{enumerate}

The pseudo code of this algorithm is given Algorithm \ref{algo:IB-KM}.
\begin{algorithm}[tp]
\label{algo:IB-KM}
\caption{IB-KM Pilot Assignment.}

\KwIn{$\left\{ {{{\bf{d}}_k}} \right\}$, $\left\{ {{{\bf{d}}'_p}} \right\}$, $\left\{ {{{\pmb \mu}' _m}} \right\}$, $\left\{ {\cal C}_{m} \right\}$, $\varepsilon$}

\KwOut{$\left\{ {{\pmb{\phi }}_k} \right\}$}

\Repeat(){$\max_{m}{\left\| {{{\pmb{\mu}}'_m} - {{\pmb{\mu }}_m}} \right\|^2_2} < \varepsilon$.}
{
    ${{\pmb{\mu }}_m}  \leftarrow  {{\pmb{\mu }}'_m},\; m = 1,\ldots,\left\lceil {K/{\tau _p}} \right\rceil$;\\
        \For{ $1 \le p \le K_{\rm p}$}
        {
            $m^* = \arg {\min _m}{\left\| {{{\bf{d}}'_p} - {{\pmb{\mu }}_m}} \right\|^2_2}$;\\
            ${{\cal C}_{m^*}} \leftarrow {{\cal C}_{m^*}} \cup \left\{ p \right\}$;\\
        }
        \For{ $1 \le m \le \left\lceil {K/{\tau _p}} \right\rceil $}
        {
          ${{\pmb{\mu }}'_m} \leftarrow \frac{1}{{\left| {{{\cal C}_m}} \right|}}\sum\limits_{{p} \in {{\cal C}_m}} {{{\bf{d}}'_p}}$;\\
        }
}
${{\pmb{\mu }}_m} \leftarrow {{\pmb{\mu }}'_m},\; m = 1,\ldots,\left\lceil {K/{\tau _p}} \right\rceil$;\\
${\cal C}_m \leftarrow \emptyset,\; m = 1,\ldots,\left\lceil {K/{\tau _p}} \right\rceil$;\\
${\cal L}_{\rm{UE}} \leftarrow \left\{1,\ldots,K \right\}$;\\
\For{ $1 \le m \le \left\lceil {K/{\tau _p}} \right\rceil $}
{
${\cal C}_m = \arg {\rm{sort}} _{k\in{\cal L}_{\rm{UE}}}{\left\| {{\rm{diag}}\left({{\bf{d}}_k}\right)  {{\bf A}_{\cdot k}} - {{\pmb{\mu }}_m}} \right\|^2_2}$;\\
/* ${\cal I} = \arg {\rm{sort}} _{i\in{\cal S}}{x_i}$ denotes the index set of the entries in $\left\{ x_i:i \in {\cal S} \right\}$, which are sorted in ascending order. */\\
${\cal C}_m \leftarrow {\left. {\cal C}_m \right|_{1, \ldots ,\tau_p}}$;\\
${\cal L}_{\rm{UE}} \leftarrow {\cal L}_{\rm{UE}}/{\cal C}_{m}$;\\
}
$\left\{ {{\pmb{\phi }}_k}:k\in {\cal C}_1  \right\}\leftarrow\left\{ {{\pmb{\phi }}_1},\ldots,{{\pmb{\phi }}_{\tau_p}}  \right\}$;\\
\For{ $2 \le m \le \left\lceil {K/{\tau _p}} \right\rceil $}
    {
        ${\cal C}'_{1}\leftarrow {\cal C}_1$;\\
        ${\cal C}'_{m} \leftarrow { {\cal C}}_{m}$;\\
        \Repeat(){${\cal C}'_{m} = \emptyset$.}
        {
            $ {{{\cal L}_{i}}}\leftarrow\emptyset :i \in {\cal C}'_{m}$;\\
            \For{ $k \in {\cal C}'_{1}$}
            {
                $i^* = \arg {\max _{i \in {\cal C}'_{m}}}{\rm{Dis}}_{ik}$;\\
                ${{{\cal L}_{i^*}}} \leftarrow {{{\cal L}_{{i^*}}}} \cup \left\{  k \right\}$;\\
            }
            \For{ $\left(i \in {\cal C}'_{m}\right) \cap \left( {{{\left| {\cal L}_i \right|}} \ne 0} \right)$}
            {
                \If{$\left| {{{\cal L}_{i}}} \right| = 1 $}
                {
                ${{\pmb{\phi }}_i}\leftarrow{{\pmb{\phi }}_{{\cal L}_{i}}}$;\\
                ${\cal C}'_{1} \leftarrow {\cal C}'_{1} / {{\cal L}_{i}}$;\\
                }
                \ElseIf{$\left| {{{\cal L}_{i}}} \right| > 1 $}
                {
                $k^* = \arg {\max _{k' \in {\cal L}_i}}{\rm{Dis}}_{ik'}$;\\
                ${{\pmb{\phi }}_i}\leftarrow{{\pmb{\phi }}_{k^*}}$;\\
                $ {\cal C}'_{1} \leftarrow {\cal C}'_{1} / \left\{  k^* \right\}$;\\
                }
                ${\cal C}'_{m} \leftarrow {\cal C}'_{m} / \left\{  i \right\}$;\\
            }
        }
    }
{\bf{final}};
\end{algorithm}

One way to view the K-means-type pilot assignment method is that it dynamically divides the network into subareas, defined by the centroids, where each pilot is only used once. From this perspective, the network is divided into cells but we stress that the rest of the processing in the network is performed in cell-free manner.
Although the IB-KM pilot assignment scheme separates the UE clusters as far as possible, it operates in the cluster level, or the centroid level.
There is still a risk that several cluster-edge UEs served by similar subsets of APs share the same pilot, like UEs sharing Pilot 3 in Fig.~\ref{fig:K-means}.
In order to further suppress the pilot contamination, we need to separate the UEs sharing the same pilot as far as possible directly at the UE level, which could not be achieved by the above K-means-type pilot assignment scheme since it is \emph{centroid-centric}.
To solve this issue, we propose the following pilot assignment scheme in a \emph{user-centric} manner.

\subsection{User-Group Pilot Assignment Scheme}\label{subsec:}
\begin{figure}[t]
\centering
\includegraphics[scale=1]{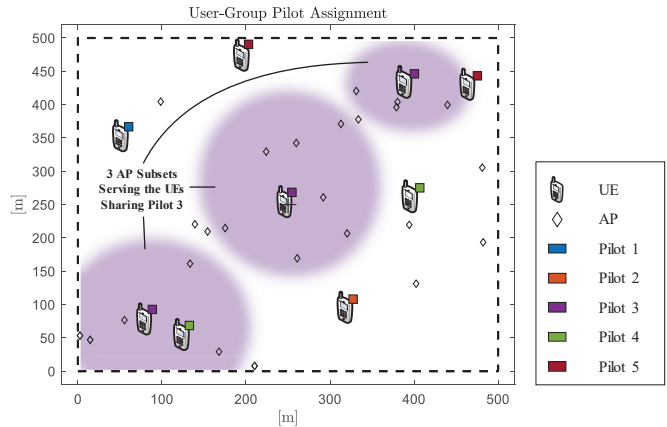}
\caption{A cell-free massive MIMO network with User-Group pilot assignment, where $9$ UEs are separated into $5$ groups. UEs in the same group share the same pilot.
\label{fig:User-Group}}
\end{figure}

The User-Group pilot assignment aims to assign mutually orthogonal pilots to the UEs served by similar subsets of APs.
The key difference from the IB-KM pilot assignment is that the User-Group pilot assignment finds the UEs having the minimum intersections of ${\cal M}_i$, ($1 \le i \le K$), then put them into the same group, and assign this group an orthogonal pilot, as shown in Fig.~\ref{fig:User-Group}.

This is reasonable since as we can see in \eqref{eq:received pilot signal in AP}, pilot contamination occurs when several UEs that share the same pilot are communicating with the same AP.
In other words, based on the proposed AP selection procedure in Section \ref{sec:ap_sele}, the fewer common serving APs the UEs have, the less pilot contamination would be caused if these UEs share the same pilot.
Based on point, our proposed User-Group pilot assignment scheme operates through the following steps.
\begin{enumerate}

  \item The CPU collects the AP selection results $\left\{ {{{\cal M}_{k}}}  \right\}$ achieved in Section \ref{sec:ap_sele} and structures a matrix ${\bf S} \in {\mathbb R}^{L \times K}$, which only keeps the \emph{strongest} serving relationships between APs and UEs indicated in $\left\{{\bf A}_{kl}\right\}$.
      Matrix ${\bf S}$ is constructed by first sorting the large-scale fading coefficient $\left\{{ \beta}_{ij}\right\}$ whose indices $\left(i,j \right)$ with ${\bf A}_{ij} = 1$, in descending order, as
            \begin{equation}\label{eq:}
            \bar {\cal A} = \left\{ {{{ \beta }_{ij}}:{\bf A}_{ij} = 1} \right\}.
            \end{equation}
            {Then, we keep the first $\left\lceil {\delta \left| {\bar {\cal A}} \right|} \right\rceil $ ${\beta}_{ij}$s as $\tilde {\cal A}$,
            where $0 < \delta  \le 1$ is a predetermined threshold determining how many serving relationships will be kept in the matrix $\bf S$, which affects the number of the groups.}
            Finally, the matrix ${\bf S}$ is constructed as
            \begin{equation}\label{eq:}
            {{\bf{S}}_{ij}} =  {\begin{cases}
            1&{{\rm{if}}\;{\beta _{ij}} \in \tilde {\cal A}}\\
            0&{{\rm{otherwise}}}
            \end{cases}}
            \end{equation}

\begin{figure*}[t!]
\centering
\includegraphics[scale=0.8]{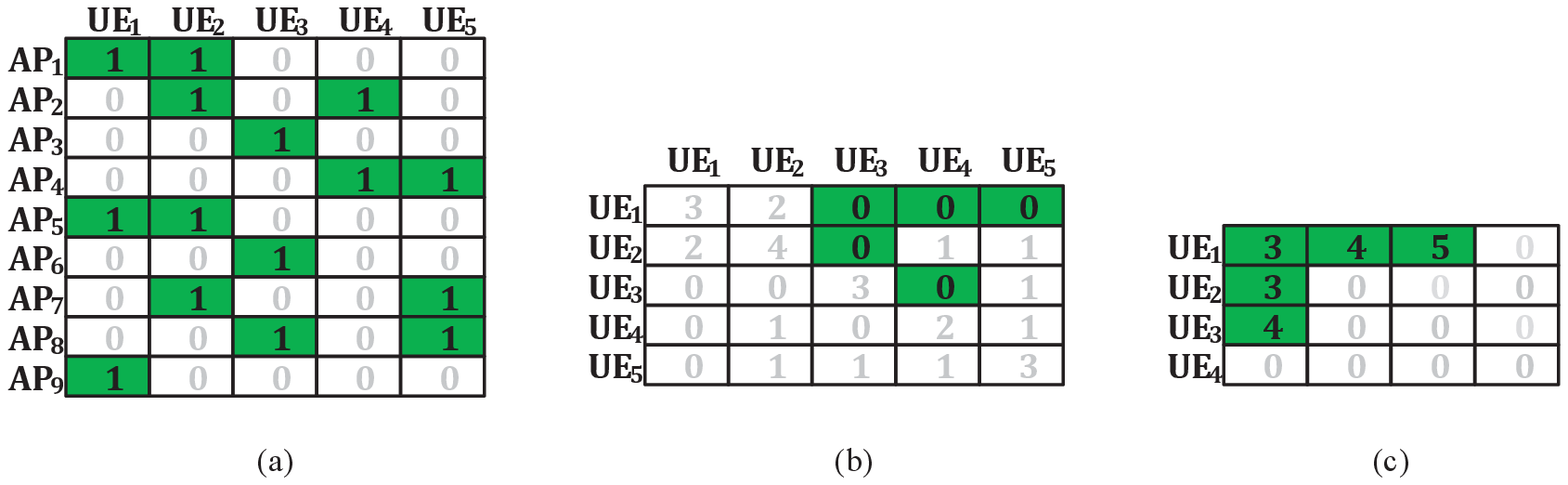}
\caption{An example of User-Group pilot assignment consisting of $5$ UEs and $9$ APs. (a) Matrix $\bf S$: AP-UE serving relationship; (b) Matrix $\bf T$: UE-UE interference relationship; (c) Matrix $\bf G$: UE-UE grouping relationship.
\label{fig:matrix}}
\end{figure*}
  \item In order to reveal the inter-user interference relationship among $K$ UEs, a matrix ${\bf T} \in {\mathbb R}^{K \times K}$ is structured as
            \begin{equation}\label{eq:}
            {\bf T} = {\bf S}^{\rm T}{\bf S}.
            \end{equation}
        The zero-valued  entries ${\bf T}_{ik}$ of the matrix ${\bf T}$ indicates that UE $k$ and UE $i$ are served by fewest common APs, i.e., ${{{\cal M'}_k} \cap {{\cal M'}_i} = \emptyset }$, where ${{\cal M'}_k}$ is the set with the nonzero entries in the $k$th column of ${\bf S}$.
        In other words, if UE $k$ wants to form a group to share a pilot, UE $i$ \emph{could} be a potential member.
        Note that ${\cal M'}_k  \subset {\cal M}_k$ is only used for user-grouping.
        Moreover, ${\bf T}$ is a symmetric matrix, thus we only focus on the entries above the main diagonal.
  \item A matrix ${\bf G} \in {\mathbb R}^{(K-1) \times (K-1)}$ is structured for the following grouping procedure, where the entries in each row of ${\bf G}$ are the column indices of the zero entries in the corresponding row of ${\bf T}$, in ascending order.
      For better elaboration, we present a simple example in Fig.~\ref{fig:matrix}, which consists of $5$ UEs and $9$ APs.
      It can be observed that the nonzero entries in the first row of matrix $\bf G$ are $\left\{3,4,5\right\}$, which are the column indices in the first row of matrix $\bf T$.

  \item We denote by ${\cal L}_{\rm{UE}} \subset \left\{ {1, \ldots ,K} \right\}$ the set of indices belonging to the UEs which are available to be selected as members of a group.
        When a UE is forming a group or has been selected as member of another UE, the index of this UE is removed from ${\cal L}_{\rm{UE}}$.

        We denote by ${\cal G}_{m_k} \subset \left\{ {1, \ldots ,K} \right\}$ the set of indices belonging to the UEs which are the members of the $m$th group, which is formed by UE $k$.
        ${\cal G}_{m_k}$ should satisfy
        \begin{equation}\label{eq:constraint}
        {{\cal M'}_i} \cap {{\cal M'}_j} = \emptyset ,\;\forall i,j \in {{\cal G}_{m_k}}.
        \end{equation}
        By denoting the set of the nonzero entries in the $k$th row of matrix ${\bf G}$ as ${\cal R}_k$, the equivalent constraint in \eqref{eq:constraint} can be  depicted as if $(i,j \in {\cal R}_k) \cap (j \notin {\cal R}_i)$, then $j \notin {\cal G}_{m_k}$.

        Note that the last UE, i.e., UE $K$ needs to be dealt with as a special case since the diagonal entries of matrix $\bf T$ are always positive.
        If UE $K$ is not selected by any group until the end of the grouping procedure, it forms a group by its own.
        The algorithm initiates with ${\cal L}_{\rm{UE}} = \left\{1,\ldots,K \right\}$ and $\left\{ {\cal G}_m = \emptyset \right\}$, where $\left| \left\{ {\cal G}_m  \right\} \right| \le K$.
\end{enumerate}
\renewcommand\arraystretch{1.5}
\begin{table}[t!]
  \centering
  \fontsize{9}{9}\selectfont
  \caption{Reference initial value of $\delta$ used in bisection method ($K = 40$). }
  \label{tab:delta}
    \begin{tabular}{|p{1.2cm}<{\centering}|p{0.9cm}<{\centering}|p{0.9cm}<{\centering}|p{0.9cm}<{\centering}|p{1.1cm}<{\centering}|}
    \hline
          &$\tau_p = 4$   &$\tau_p = 6$  &$\tau_p = 8$   &$\tau_p = 10$     \cr\hline
        $L = 121$    &$0.24$   &$0.27$  &$0.30$   &$0.32$  \cr\hline

        $L = 196$   &$0.21$   &$0.23$  &$0.25$   &$0.27$  \cr\hline
    \end{tabular}
\end{table}

The grouping procedure separates the $K$ UEs into $M$ disjoint groups for a given threshold $\delta$, thus we need to adjust $\delta$ to achieve $M = \tau_p$.
Bisection method could be applied on $\delta$ to obtain the desired $M = \tau_p$ dynamically, since $\left|{\cal M}_k\right|$, $k = 1,\ldots,K$ reduces (i.e., the circle in Fig.~\ref{fig:User-Group} shrinks) as the threshold $\delta$ reduces, which increase the chance of ${\cal M'}_i \cap {\cal M'}_k = \emptyset, \forall i,k \in \left\{{ 1,\ldots,K }\right\}$.
We give some reference initial value of $\delta$ used in bisection method with several setups in Table~\ref{tab:delta}.
The pseudo code of this algorithm is Algorithm \ref{algo:user group}.

{\subsection{Online Complexity Analysis}}
{
The random pilot assignment operates over $K$ UEs where each UE randomly chooses a pilot, hence, the corresponding complexity is $\mathcal{O}(K)$.
IB-KM operates in two steps, i.e., locating $\left\lceil {K/{\tau _p}} \right\rceil$ centroids and assigning $K$ UEs to these clusters.
Since the locations of the centroids are determined by the geographic locations of the APs, which is a-priori known at the CPU, the first step of IB-KM could be finished offline before the transmission commences, and can therefore be neglected when counting the online complexity.
The complexity of IB-KM depends on the second step, in which each UE selects its centroid based on the distances between it and all $\left\lceil {K/{\tau _p}} \right\rceil$ centroids. Each UE in ${\cal C}_1$ finds one unique UE from each of the other $\left\lceil {K/{\tau _p}} \right\rceil - 1$ clusters to share its pilot.
Therefore, the complexity of the IB-KM becomes $\mathcal{O}\left(K^2/\tau_p + \tau_p^2\left(\left\lceil {K/{\tau _p}} \right\rceil - 1\right)\right)$.
User-Group requires computation of the matrices $\bf S$,  $\bf T$, and $\bf G$.
Note that only the entries above the main diagonal of matrix $\bf T$ are exploited to construct the matrix $\bf G$.
Therefore, the complexity of User-Group becomes $\mathcal{O}\left(KL + K^2L + K/2 \right)$.
For the considered massive access cell-free massive MIMO system, the number of pilots is far smaller than the number of APs and UEs, i.e., $L \approx K \gg \tau_p$ is satisfied. Thus, the IB-KM scheme has a much more attractive complexity scaling than the User-Group scheme.
}
\begin{algorithm}[tp]
\label{algo:user group}
\caption{User-Group Pilot Assignment Algorithm.}

\KwIn{$\left\{ {\cal R}_k \right\}$, ${\cal{L}}_{\rm{UE}}$, $\left\{ {\cal G}_m \right\}$, $\tau_p$, $\delta$, $\delta_{\rm {min}}$, $\delta_{\rm {max}}$}

\KwOut{$\left\{ {{{\cal G}_m}:m \in \left\{1, \ldots ,K\right\}} \right\}$}

\Repeat(){\bf{break}}
{
    ${\mathcal F} \leftarrow 1$;\\
    $m \leftarrow 0$;\\
    \Repeat(){${\cal{L}}_{\rm{UE}} =\emptyset$}
    {
         $m \leftarrow m +1 $;\\
         {$i^* \leftarrow  {\cal L}_{\rm{UE}} \left(1\right)$};\\
           \If{$i^* = K$}
            {
                ${\mathcal G}_{m} \leftarrow \left\{ {K} \right\}$;\\
                ${\mathcal F} \leftarrow 0$;\\
            }
         ${\mathcal G}_{m} \leftarrow \left\{ {i^*} \right\}$;\\
         ${{\cal L}_{{\rm{UE}}}} \leftarrow {{\cal L}_{{\rm{UE}}}}/\left\{ {{i^*}} \right\}$;\\
         ${{\cal R}_{k}} \leftarrow {{\cal R}_{k}}/\left\{ {i^*} \right\},\forall k \in {{\cal L}_{{\rm{UE}}}}$;\\
        \Repeat(){${\cal{R}}_{i^*} =\emptyset$}
        {
            {$j^* \leftarrow  {\cal R}_{i^*} \left(1\right)$};\\
               \If{$j^* = K$}
            {
                ${\mathcal F} \leftarrow 0$;\\
            }
            ${\mathcal G}_{m} \leftarrow {\mathcal G}_{m} \cup \left\{ {j^*} \right\}$;\\
            ${\cal R}_{i^*} \leftarrow {\cal R}_{i^*} \cap {\cal R}_{j^*}$;\\
            ${{\cal L}_{{\rm{UE}}}} \leftarrow {{\cal L}_{{\rm{UE}}}}/\left\{ {{j^*}} \right\}$;\\
            ${{\cal R}_{k}} \leftarrow {{\cal R}_{k}}/\left\{ {j^*} \right\},\forall k \in {{\cal L}_{{\rm{UE}}}}$;\\
        }
    }
    \If{${\mathcal F} = 1$}
            {
                $m \leftarrow m +1 $;\\
                ${\mathcal G}_{m} \leftarrow \left\{ {K} \right\}$;\\
            }
    $M \leftarrow m$;\\
    \If{$M = \tau_p$}
    {
        \For{$1 \le m \le M$}
        {
            ${{\pmb{\phi }}_k} \leftarrow {{\pmb{\phi }}_m}, k\in {\cal G}_m$;\\
        }
        {\bf{break}};\\
    }
    \ElseIf{$G < \tau_p$}
    {
        $\delta_{\rm {min}} \leftarrow \delta$;\\
    }
    \Else
    {
        $\delta_{\rm {max}} \leftarrow \delta$;\\
    }
    $\delta  \leftarrow \left( {\delta_{\rm {min}}  + \delta_{\rm {max}} } \right)/2$;\\
}
{\bf{final}};
\end{algorithm}

\section{Scalable Fractional Power Control}\label{sec:power control}
In practical implementations, a power control policy with scalability and low complexity is needed.
Inspired by \cite{nikbakht2019uplink}, we propose a scalable fractional power control policy for data transmission, which locally minimizes the variance of the large-scale signal-interference ration (SIR), i.e.,
\begin{equation}\label{eq:SIR}
{\rm SIR}_k = \frac{{{p_k}{{\left( {\sum\limits_{l \in {{\cal M}_k}} {{\beta _{kl}}} } \right)}^2}}}{{\sum\limits_{i = 1,i \ne k}^K {{p_i}\sum\limits_{l \in {{\cal M}_k}} {{\beta _{kl}}} {\beta _{il}}} }}.
\end{equation}
Note that \eqref{eq:SIR} is derived from \cite[Eq. (18)]{nikbakht2019uplink}, where the local-average desired signal power only consists of the large-scale fading coefficients of the APs selected by UE $k$.
\begin{lemm}\label{prop:}
The data transmission power $p_k$ for UE $k$ is
\begin{equation}\label{eq:power control}
{p_k} = \frac{\eta }{{{{\left( {\sum\limits_{l \in {{\cal M}_k}} {{\beta _{kl}}} } \right)}^\theta }}}\bar p,
\end{equation}
where the scaling $\eta$ is given by
\begin{equation}\label{eq:}
\eta  = {\min _{1\le i \le K}}{\left( {\sum\limits_{l \in {{\cal M}_i}} {{\beta _{il}}} } \right)^\theta },
\end{equation}
and the parameter $\theta \in \left[{0,1}\right]$ indicates the extent to which the range of the received powers is compressed.
{Smaller values of $\theta$ favor the average SIR and  larger values} of $\theta$ promote more fairness.
\end{lemm}
\begin{IEEEproof}
It follows the similar approach as in \cite[App. A]{nikbakht2019uplink}, but for the local-average desired signal power as ${\sum\limits_{l \in {{\cal M}_k}} {{\beta _{kl}}} }$.
\end{IEEEproof}

\section{Numerical evaluation}\label{sec:numerical results}

In this section, we evaluate the proposed massive access framework and validate the closed-form SE expressions provided in Lemma \ref{lemm:MR}.
We consider a setup with $L = 100$ APs and where $K$ UEs are independently and uniformly distributed in a $0.5 \textrm{ km} \times 0.5$ km square coverage area.
The APs could be deployed on a square grid or randomly; all APs are equipped with half-wavelength-spaced uniform linear arrays with $N =4$ antennas.
We apply the wrap-around technique to approximate an infinitely large network with $1600$ antennas/km$^2$.

We apply the access and AP selection algorithm proposed in Section \ref{sec:ap_sele} when the UEs access the network. Pilots are assigned according to the pilot assignment schemes described in Section \ref{sec:pilo_assi}.
{The 3GPP Urban Microcell model in \cite[Tab. B.1.2.1-1]{access2010further} is used to compute the large-scale propagation conditions, such as pathloss and shadow fading.
Beyond that, we adopt the same system setup parameters as in \cite{bjornson2019making}, where the maximum UE transmit power is $\bar p = 100$ mW, the bandwidth is $20$ MHz, and the coherence blocks contain $\tau_c = 200$ channel uses, which could be achieved by $2$ ms coherence time and $100$ kHz coherence bandwidth (there are many possible combinations).
Unless specified, $\tau_p = 10$ channel uses are utilized for uplink pilots and the remainder is used for downlink data.}
Each UE transmits the pilot signal with full power $p_k = \bar p$, and exploits the power control during the uplink data transmission.
{In the figures, we use ``User-Group", ``IB-KM", ``GB-KM", ``Random", ``Switch", and ``Scalable" to denote the User-Group pilot assignment, IB-KM pilot assignment, GB-KM pilot assignment, random pilot assignment, random pilot switch, and the initial access and pilot assignment scheme proposed in \cite{bjornson2019scalable}, respectively.}

\begin{figure}[t]
\centering
\includegraphics[scale=0.6]{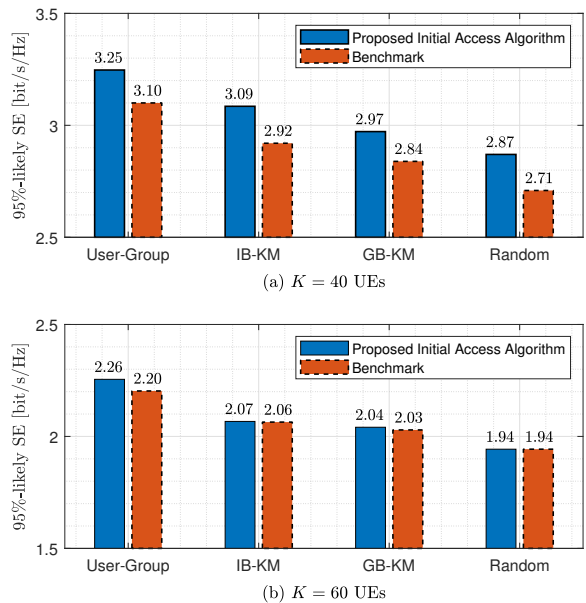}
\caption{95\%-likely SE with different combinations of initial access algorithms, pilot assignment schemes, and numbers of UEs (LP-MMSE combining, P-LSFD, $\theta = 1$).
\label{fig:95SE_MMSE_K_APselection}}
\end{figure}
In order to evaluate the performance of our proposed scalable initial access algorithm, we first consider a benchmark algorithm where each AP serves the $\tau_p$ UEs with the strongest channel conditions.
To mimic a practical scenario, we consider the random deployment of APs in this comparison.
Fig.~\ref{fig:95SE_MMSE_K_APselection} compares the proposed initial access algorithm and benchmark algorithm in 95\%-likely SE with $K=40$ and $K=60$ UEs.
The first observation is that the proposed initial access algorithm outperforms the benchmark algorithm, for all the four considered pilot assignment schemes and in both setups ($K=40$ and $K=60$).
The reason is that the competition mechanism in the proposed initial access algorithm allows each UE to be served by as many APs as possible, at the precondition of satisfying Assumption \ref{assu:1}.
When comparing Fig.~\ref{fig:95SE_MMSE_K_APselection}(a) and Fig.~\ref{fig:95SE_MMSE_K_APselection}(b), we notice that the advantage of the proposed competition mechanism gets less prominent when the number of UEs gets larger, and each UE can only get limited service for both cases due to the high UE density.
When the APs are deployed on a square grid, the advantage of the competition mechanism compared with the benchmark becomes limited; however, the sum SE of the network is improved by the reduction of the low-rate UEs.
Therefore, we apply the grid deployment for the APs in the following numerical results.

\begin{figure}[t]
\centering
\includegraphics[scale=0.6]{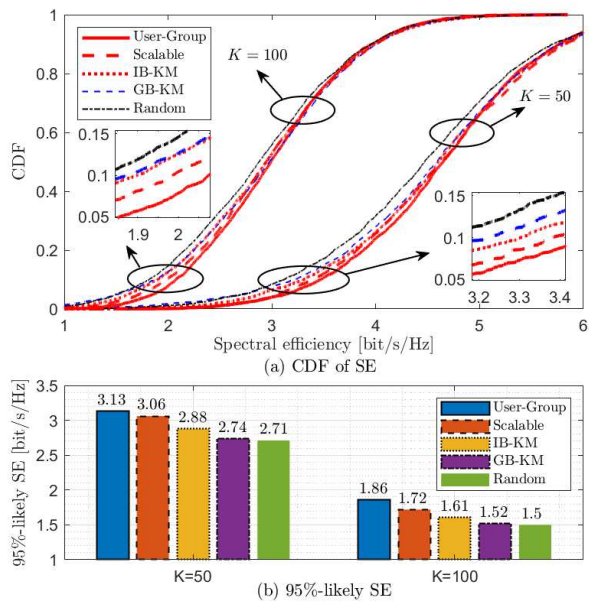}
\caption{SE per UE with different combinations of pilot assignment schemes and numbers of UEs (LP-MMSE combining, P-LSFD, $\theta = 1$).
\label{fig:SE_CDF_MMSE_K}}
\end{figure}
Fig.~\ref{fig:SE_CDF_MMSE_K} depicts the cumulative distribution function (CDF) of the SE per UE when the LP-MMSE combining and the proposed P-LSFD in \eqref{eq:partial LSFD} are applied.
{We compare the proposed User-Group pilot assignment scheme and IB-KM pilot assignment scheme with three benchmarks, which are the GB-KM pilot assignment, random pilot assignment, and the scheme proposed in \cite{bjornson2019scalable}, respectively.
In both setups ($k=50$ and $K=100$), It can be observed from Fig.~\ref{fig:SE_CDF_MMSE_K}(a) that the User-Group and IB-KM schemes achieve better performance than the benchmarks except the Scalable scheme, which provides better performance than IB-KM scheme while falls behind User-Group scheme.
More specifically, Fig.~\ref{fig:SE_CDF_MMSE_K}(b) shows that compared with GB-KM, User-Group achieves $14.2\%$ and $22.4\%$ improvement in 95\%-likely SE for the cases of $K = 50$ and $K = 100$, respectively; IB-KM achieves  $5.1\%$ and $5.9\%$ improvement in 95\%-likely SE in these two setups, respectively.
Moreover, compared with Scalable, User-Group achieves $2.3\%$ and $8.1\%$ improvement in 95\%-likely SE in these two setups, respectively.
When comparing the two setups, we observe that the large density of UEs benefits the improvement of the proposed User-Group and IB-KM schemes.
The reason behind this is the User-Group and IB-KM schemes are dedicated to suppressing the inter-user interference, which is much stronger in a massive access scenario.}

\begin{figure}[h]
\centering
\includegraphics[scale=0.6]{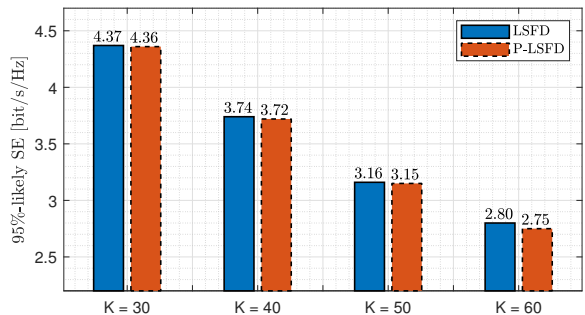}
\caption{95\%-likely SE with different combinations of data decoding strategies and numbers of UEs (LP-MMSE combining, User-Group pilot assignment, $\theta = 1$).
\label{fig:SE_MMSE_LSFD}}
\end{figure}

The performance of the proposed P-LSFD is evaluated through Fig.~\ref{fig:SE_MMSE_LSFD}.
Since we focus on the performance loss of P-LSFD comparing with LSFD, we consider the 95\%-likely SE of the User-Group scheme with LP-MMSE combining for the setups of $K=30,40,50,60$ UEs, respectively.
Among these four setups, we notice that the P-LSFD achieves roughly the same 95\%-likely SE.
{The performance loss of P-LSFD compared to LSFD increases as the number of UEs increases, due to the fact that each AP can only serve a maximum of number of UEs, thus an increasing number of UEs leads to fewer serving APs per UE.}
However, the largest performance loss in this comparison is only $1.8\%$ when $K = 60$, which implies that the scalability on P-LSFD can be achieved with a negligible performance loss.

\begin{figure}[t]
\centering
\includegraphics[scale=0.6]{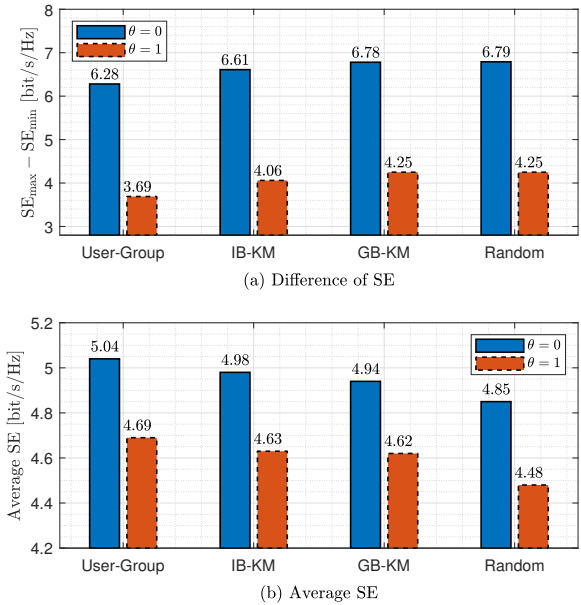}
\caption{Fairness and average SE with different combinations of pilot assignment schemes and power control parameters, (LP-MMSE combining, P-LSFD, $K =50$).
\label{fig:SE_MMSE_theta}}
\end{figure}

{Fig.~\ref{fig:SE_MMSE_theta} illustrates the proposed scalable fractional power control policy in fairness and average SE for the setup of $K=50$, respectively.}
Note that the scalable fractional power control policy comprises the so-called \emph{equal power allocation} by letting $\theta =0$. Furthermore, the fairness is measured by the difference between the maximum and minimum values of the SE, i.e., ${\rm{SE}}_{\rm{max}} - {\rm{SE}}_{\rm{min}}$.
{It can be observed from Fig.~\ref{fig:SE_CDF_MMSE_K}(a) that larger values of $\theta$ promotes more fairness among the UEs.}
Since for one UE, the disadvantage in the large-scale fading coefficients between its serving APs will be compensated with the transmission power.
According to \eqref{eq:power control}, the larger the value of $\theta$ is, the lager the compensation is.
Another observation is that ${\rm{SE}}_{\rm{max}} - {\rm{SE}}_{\rm{min}}$ is insensitive with respect to the number of UEs.
{Moreover, Fig.~\ref{fig:SE_CDF_MMSE_K}(b) shows that smaller values of $\theta$ improves the average SE} since the transmission power of each UE in the network approaches to the maximum power $\bar p$ as $\theta \to 0$.
It is clear to see that the average SE decreases as the number of UEs increases, because the strong inter-user interference caused by the high density of UEs accessing the network with limited pilots.
When comparing the four pilot assignment schemes, it is clear that the proposed User-Group and IB-KM schemes outperform the GB-KM and random methods in both terms of UEs fairness and average SE.

\begin{figure}[h]
\centering
\includegraphics[scale=0.6]{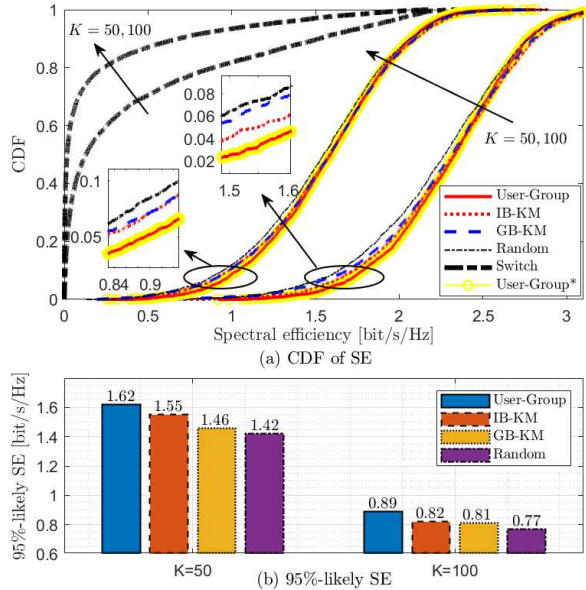}
\caption{SE per UE with different combinations of pilot assignment schemes and numbers of UEs (MR combining, P-LSFD, $\theta = 0$).
\label{fig:SE_CDF_MR_K}}
\end{figure}

\begin{figure}[h]
\centering
\includegraphics[scale=0.6]{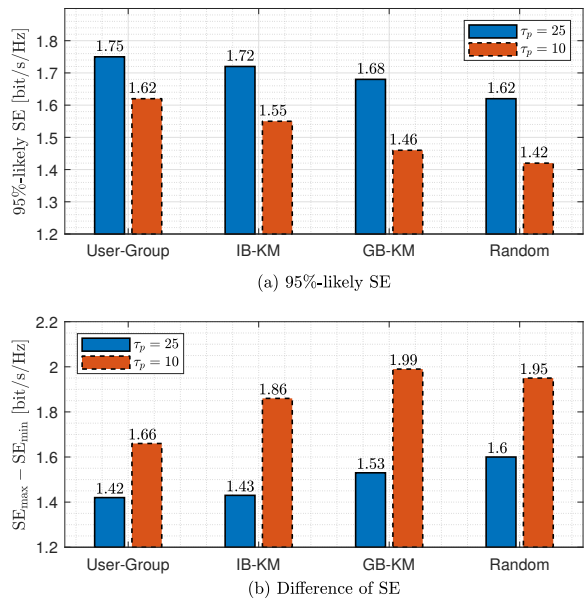}
\caption{95\%-likely SE with different combinations of pilot assignment schemes and numbers of pilots (MR combining, P-LSFD, $K = 50$, $\theta = 0$).
\label{fig:SE_MR_tau}}
\end{figure}
Since we have demonstrated the proposed scalable P-LSFD strategy, fractional power control policy, and pilot assignment schemes perform well with LP-MMSE combining, {the following results focus on the performance with MR combining, the impact of the number of the pilots, and the tightness of the closed-form SE expression provided in Lemma \ref{lemm:MR}, which are marked with ``(User-Group*)" in Fig.~\ref{fig:SE_CDF_MR_K}.}
The curve ``Switch" is plotted based on the analytical results obtained from Corollary \ref{coro:random SINR}.
The first observation is that the performance gap between the random pilot switching and other pilot assignment schemes is large.
The reason is that the strong mutual interference only occasionally occurs when two pilot-sharing UEs are close to each other, when the random pilot assignment is used; however, in random pilot switching, all UEs are subject to strong pilot contamination part of the time. Each UE switches its pilot sequence randomly over blocks and consequently nearby UEs are possibly sharing the same pilots.
It is significant that the analytical results from Lemma \ref{lemm:MR} achieve remarkable tightness compared with the simulation results.
Compared with Fig.~\ref{fig:SE_CDF_MMSE_K}, it can be observed that LP-MMSE combining achieves much better SE performance than the one of MR combining due to the advanced signal processing.
{Moreover, Fig.~\ref{fig:SE_MR_tau} demonstrates the impact of the number of pilots, in which we can observe the improved user fairness and the 95\%-likely SE with more pilot resources, i.e., in the setup of $\tau_p = 25$.
Since the bottleneck of the performance improvement is the strong pilot contamination caused by the pilot resource limitation, every UE in the system could obtain better service when this limitation is alleviated. }

\section{Conclusion}\label{sec:conclusion}
When scalability is considered in the uplink of cell-free massive MIMO systems, structured massive access provides a new opportunity to achieve higher SE to more users.
The bottleneck of structured massive access, i.e., the pilot contamination caused by pilot sharing, was much relieved by the proposed scalable initial access algorithm, User-Group, and IB-KM pilot assignment schemes in our framework.
The SE with LP-MMSE and MR combining was considered to evaluate this framework, where the user density and fairness among UEs were taken into account.
Two new closed-form SE expressions with MR combining were derived.
Although the analysis focused on the uplink, similar results could be expected in the downlink due to the channel reciprocity. {Since the proposed schemes make use of the geometry, they can also be applied in cases with multi-antenna UEs, but the exact details are left for future work. They can also be applied in a wider class of fading distributions than Rayleigh fading.}

The simulation results show that our proposed framework performs well compared to the state-of-the-art.
Specifically, our proposed initial access algorithm enables each UE to be served by as many APs as possible at the precondition of scalability.
Compared with the optimal LSFD, the 95\%-likely SE reduces as the user density increases when using the proposed P-LSFD, but it is marginal (1.7\% when $K = 60$) and thus an acceptable price of scalability.
{By actively suppressing the inter-user interference, the proposed User-Group and IB-KM pilot schemes offer $22.4\%$ and $5.9\%$ improvement in 95\%-likely SE, compared to GB-KM scheme ($K = 100$), respectively; User-Group scheme offers $8.1\%$ improvement in 95\%-likely SE compared to Scalable scheme ($K = 100$).}
Moreover, the User-Group algorithm is performed in a user-centric manner, which makes it capable of offering higher SE performance than IB-KM, especially when the scenario gets dense.
Finally, the proposed scalable fractional power control provides the trade-off of the fairness among the users and the average SE.

This paper provides a feasible solution for structured massive access in cell-free massive MIMO systems.
Although we focus on the SE performance with user density and fairness into account, it is straightforward to generalize the framework to also study other important factors, such as energy efficiency, hardware impairment, limited fronthaul capacity, etc.

\bibliographystyle{IEEEtran}
\bibliography{IEEEabrv,SCh_MA_CFmMIMO}

\end{document}